\documentclass{aa} 

\usepackage[normalem]{ulem}

\usepackage{afterpage}

\usepackage{graphicx}
\usepackage{makecell}
\usepackage{floatrow}

\usepackage{txfonts}
\usepackage{longtable}
\usepackage{comment}
\usepackage{float}

\usepackage{dblfloatfix}

\usepackage[colorlinks=true,linkcolor=blue,citecolor=blue,urlcolor=blue]{hyperref}

\begin{document}

\title{Differential rotation of solar $\alpha$ sunspots and implications for stellar light curves}

\author{Emily Joe Lößnitz\inst{1,2}
    \and
    Alexander G.M. Pietrow\thanks{For further information, data requests, or correspondence contact A. G. M. Pietrow (apietrow@aip.de)}\fnmsep\inst{1}
    \and
    Hritam Chakraborty\inst{3}
    \and
    Meetu Verma\inst{1}
    \and
    Ioannis Kontogiannis\inst{1,4,5}
    \and
    Horst Balthasar\inst{1}
    \and
    Carsten Denker\inst{1}
    \and
    Monika Lendl\inst{3}
    }

   \institute{Leibniz-Institut für Astrophysik Potsdam (AIP), An der Sternwarte 16, 14482 Potsdam, Germany
   \and
   Universität Potsdam, Institut für Physik und Astronomie, Karl-Liebknecht-Straße 24/25, 14476 Potsdam, Germany
   \and
   Geneva Observatory, University of Geneva, Chemin Pegasi 51, 1290 Versoix, Switzerland
   \and
   Eidgen\"ossische Technische Hochschule Z\"urich, Wolfgang-Pauli-Str. 27, 8093, Z\"urich, Switzerland
   \and
    Istituto Ricerche Solari Aldo e Cele Dacc\`o (IRSOL), Via Patocchi 57,  6605 Locarno, Switzerland
   }

\date{Draft: compiled on \today}

\abstract
{Differential rotation is a key driver of magnetic activity and dynamo processes in the Sun and other stars, especially as the rate differs across the solar layers, and also in active regions.}
{We aim to accurately quantify the velocity at which round $\alpha$ spots traverse the solar disk as a function of their latitude, and compare these rates to those of the quiet Sun and other sunspot types. We then extend this work to other stars and investigate how differential rotation affects the modulation of stellar light curves by introducing a generalized stellar differential rotation law.}
{We manually identified and tracked 105 $\alpha$ sunspots in the 6173~\AA\ continuum using the Helioseismic and Magnetic Imager (HMI) aboard the Solar Dynamics Observatory (SDO). We measured the angular velocities of each spot through center-of-mass and geometric ellipse-fitting methods to derive a differential rotation law for round $\alpha$ sunspots.}
{Using over a decade of HMI data, we derived a differential rotation law for $\alpha$ sunspots. Compared to previous measurements, we find that $\alpha$ sunspots rotate 1.56\% faster than the surrounding quiet Sun, but 1.35\% slower than the average sunspot population. This supports the hypothesis that the depth at which flux tubes are anchored influences sunspot motions across the solar disk. We extend this analysis to other stars by introducing a scaling law based on the rotation rates of these stars. This scaling law is implemented with the Stellar Activity Grid for Exoplanets (SAGE) code to illustrate how differential rotation alters the photometric modulation of active stars. 
}
{Our findings emphasize the necessity of considering differential rotation effects when modeling stellar activity and exoplanet transit signatures.}


 \maketitle

\section{Introduction}

Rotation is a fundamental property of stars. Accurate measurements of this property enable several astrophysical phenomena to be investigated, including the strength, morphology, and lifetime of magnetic fields, the characteristics of magnetic dynamos, and the outflow rates of stellar material. The rotation rate varies according to spectral type, with an overall trend of slowing as the effective temperature decreases \citep{Barnes2016, Gilhool2019}. 

The Sun exhibits differential rotation, a phenomenon in which different regions of a rotating body exhibit varying angular velocities. This effect is caused by shear in the solar atmosphere, leading to faster rotation at the equator that decreases toward the poles. \citet{Carrington1863} first discovered and measured this phenomenon on the Sun in 1863 when he observed that photospheric features rotate more rapidly at the equator than at higher latitudes.

\begin{figure}[t]
    \centering
    \vspace{-0.2cm}
    \includegraphics[width=0.99\columnwidth]{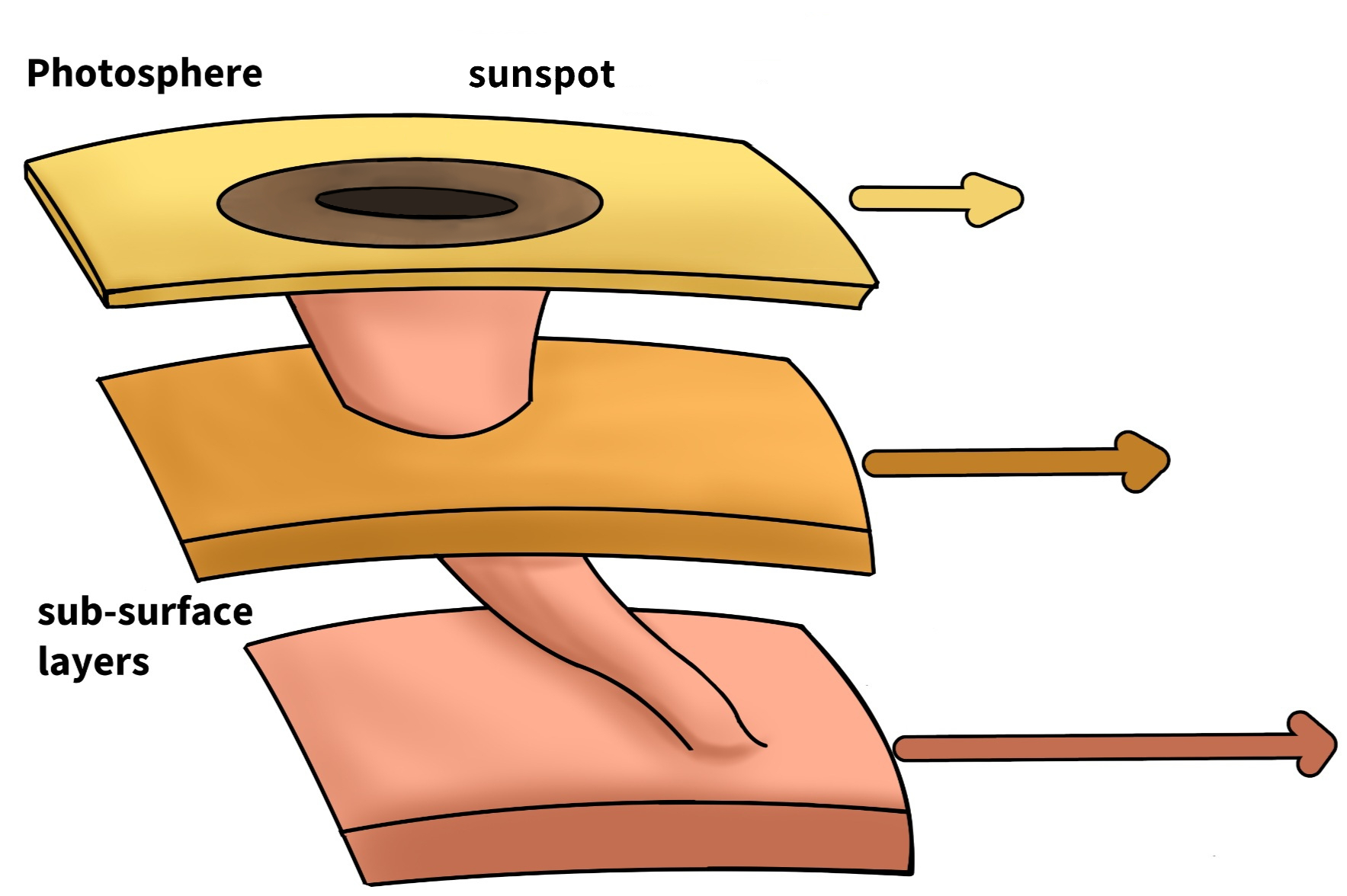}
    \caption{Cartoon depicting a sunspot being anchored in deeper subsurface layers of the convection layer and dragged along with it. The rotation rate (indicated by the respective length of the arrows) varies with depth.}
    \label{fig:deep_layer_sunspot_connection}
\end{figure}

Since Carrington's discovery, different methods of measuring solar differential rotation have been developed such as sunspot and magnetic feature tracking methods \citep[e.g.,][]{Snodgrass1983, NewtonNunn1951, Ward1966, Howard1984, Balthasar1986, Kutsenko2023}, Doppler shift measurements \citep[e.g.,][]{Snodgrass1984, HowardAdkins1983}, and helioseismology \citep[e.g.,][]{Benevolenskaya1999, Howe2000, SchouHowe2002, Korhonen2021}. Similar studies in the higher-lying chromosphere and corona reveal distinct differential rotation rates compared to the photosphere that also vary over the solar cycle \citep[e.g.,][]{Wan2023, Li2024, Mishra2024, Rao2024}.
Many of these methods rely on tracking sunspots or other features on the solar surface. However, Doppler shift measurements have shown that the rotation rate of sunspots can differ from those of the surrounding photosphere by up to 4\% \citep[e.g.,][]{Foukal1979, Balthasar1986}. This ``differential rotation difference'' occurs due to a coupling of the sunspot flux tube to the underlying convection zone. There, the entire structure is dragged ahead of the overlying material by the faster-rotating, lower-lying layer (see Fig.~\ref{fig:deep_layer_sunspot_connection}). Therefore, the rotation rate of sunspots follows interior rotation rates at certain layers. The mean rotation rate of young spots corresponds to the internal rate at 0.93 solar radii \citep{Beck2000}. As the spot ages, the flux tube decouples and rises, slowing down the spot \citep{Balthasar1982, Solanki2003}.

The rotation of stars can be measured by using high-precision photometry from, for example, the Kepler satellite \citep{2010Kepler}, the CHaracterizing ExOPlanet Satellite \citep[CHEOPS,][]{Cheops2013}, and the Transiting Exoplanet Survey Satellite \citep[TESS,][]{Tess2015}. In this case, active regions on the stellar surface modulate the light curve when they rotate onto and across the solar disk \citep{Hritam2024}. These regions occur at different latitudes and can be affected by differential rotation \citep{Reinhold2015}.

The concept of differential rotation in stars was first discussed by \citet{Goldreich1967} and later confirmed using Zeeman-Doppler imaging \citep{Donati1997}, Fourier analysis of spectral lines \citep[e.g.,][]{Reiners2002}, the analysis of photometric light curve modulation \citep[e.g.,][]{Reinhold2013}, the Rossiter-McLaughlin effect \citep[e.g.,][]{Cegla2016}, and tracking of spot eclipses during repeated planetary transit observations \citep[e.g.,][]{Netto2020}. These studies allow for a more robust understanding of the shear and show a difference in the strength and prevalence of differential rotation in different stellar types. The shearing factor is weakest in G, K, and M stars, then increases significantly for F stars, and is strongest in A and B stars that possess an outer convective zone \citep{Reiners2007, Balona2016, Zaleski2020}. For G-type stars a similar shear value was found for most stars independent of their equatorial rotation, while hotter stars show more scatter \citep{Reinhold2013}. On the other hand, younger, faster-rotating stars have been found to display less differential rotation \citep{Kitchatinov1995, Reiners2007}. These trends suggest that while the degree of shear varies, differential rotation is a common feature across stars, making accurate differential rotation measurements on the Sun a valuable baseline for broader stellar application, provided appropriate scaling is taken into account.

In this work, we focus on inferring the differential rotation rate of round unipolar spots, also known as $\alpha$ spots according to the \citet{Hale1919} classification. These spots primarily consist of older decaying spots but can also include younger emerging spots. We focus on these spots because they are the most commonly simulated type of spots in detailed magnetohydrodynamic simulations of sunspot structure and dynamics \citep[e.g.,][]{Rempel2009, Schmassmann2021} as well as in stellar models \citep[e.g.,][]{Soap2012, Cauley2018, Petit2024}. We compare these values with photospheric rotation rates as well as subsurface rotation rates from helioseismology and propose a way to extrapolate this effect to other stars, which in turn is used to study the impact of this effect on photometric exoplanet retrievals. By modeling the effect of differential rotation on light-curve modulation, we believe that we can constrain the strength or shear of differential rotation in stars other than the Sun.

\section{Observations and methods}\label{observations}

In this section, we describe the selection and tracking of sunspots and our approach to deriving a rotation law. Finally, we suggest a method of extrapolating this law to other stars and describe how it is implemented in a toy model.

\subsection{Sunspot selection and tracking}

For this study, 105 sunspots were manually identified (see Table \ref{tab:all_sunspots}) using the solar monitor \citep{Gallagher2002} and then processed with the {\texttt{SpotiPy}} Python package.\footnote{\url{https://github.com/Emily-Joe/SpotiPy}} A sunspot was considered well suited for this study if it met two main criteria: 
\begin{enumerate}
    \item It should be unipolar, preferably round, and should not experience drastic changes during its disk crossing such as splitting, rapid expansion, or decay, or the formation of other active regions nearby.
    \item It must be visible and stable for at least 10 days, which is slightly more than half the time a feature takes to traverse the solar disk.
\end{enumerate}
The sample therefore consists almost exclusively of sunspots that fit the $\alpha$ category of the \citet{Hale1919} classification. All $\alpha$ sunspots analyzed in this study were already fully formed when they first appeared at the eastern limb, indicating that they were at least one day old. None of them formed during the disk passage. This finding aligns with earlier studies by \citet{NewtonNunn1951} that suggest that $\alpha$ spots are old, recurring spots.

\begin{figure*}[t]
    \centering
    \floatbox[{\capbeside\thisfloatsetup{capbesideposition={right,top},capbesidewidth=5cm}}]{figure}[\FBwidth]
    {\caption{Grading system for the considered $\alpha$ sunspot is based on its stability during its transit, and whether or not it has companions, and how it evolves. The highest grade of 3 is given to exemplary $\alpha$ spots that are visible for an entire crossing of the solar disk, which are isolated without other features nearby, and which stay in a mostly constant shape. The examples for each sunspot-grade category were created by overlaying images from each time series.}\label{fig:sunspot_scores}}
    {\includegraphics[width=13cm]{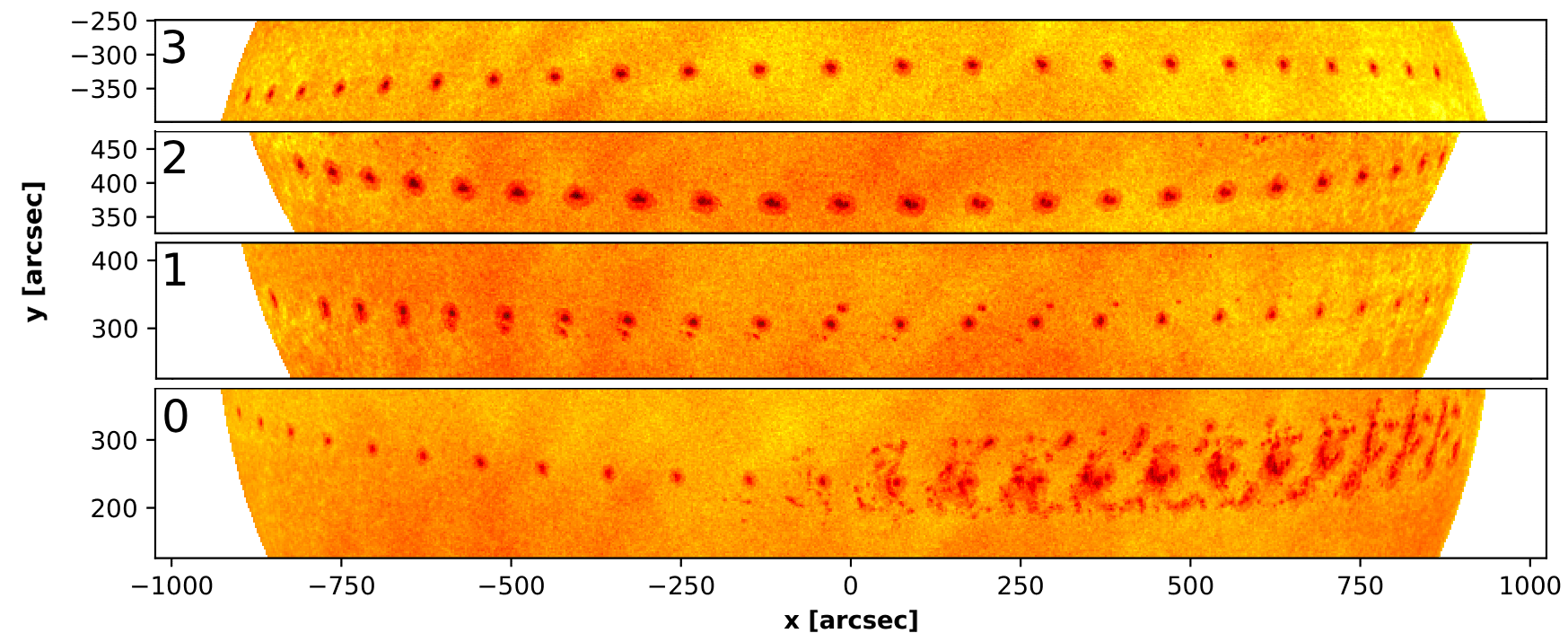}}
\end{figure*}

These criteria provide a suitably large sample of the same sunspot type, eliminating biases introduced by the evolution of complex active regions. However, an additional weighting factor was assigned to each spot crossing that represents its overall resemblance to a model $\alpha$ sunspot. This factor was based on a manually assigned grade between 0 and 3. The highest grade was assigned to spots that stay relatively unchanged during their crossing and do not have secondary spots or pores close by. In our sample, 23 sunspots received this grade. Spots with a stable crossing but irregular shape received a grade of 2 (40 sunspots in our sample), and a grade of 1 was assigned to spots with nearby spots or pores, or if they change strongly after 10 or more days (35 sunspots). Spots that were technically considered to be $\alpha$ sunspots but that did not fall within our criteria received a grade 0 and were therefore not considered in the analysis of the rotation parameters. This applied to seven sunspots in our sample, which would bring the total of sunspots used for analysis down to 98. Examples of each grade are shown in Fig.~\ref{fig:sunspot_scores}.

For every suitable sunspot, a range of dates was defined over which it passed the solar disk as well as an observational cadence. Then the \ion{Fe}{I} 6173~\AA\ line continuum as observed by the Helioseismic and Magnetic Imager \citep[HMI,][]{Scherrer2012} was downloaded from the Solar Dynamics Observatory \citep[SDO, ][]{Pesnell} database. For each time step, an intensity-calibrated full-disk image without limb darkening was downloaded.

The limb-darkening removal is part of the standard pipeline, and the data products were created by dividing by a radially fit fifth-order polynomial in terms of $\log(\mu)$ with
\begin{align}
   \mu = \sqrt{1 - r/r_\odot},
\end{align}
where $\mu$ is also defined as the cosine of the heliocentric angle, so that $\mu$ is zero at the solar limb and unity at the center (Charles Baldwin, private communication, January 18, 2024). 

Historically, irregular cadences of around 24 hours were used for sunspot analysis as this corresponded to daily observations \citep[e.g.,][]{BalthasarWoehl1980, Howard1984, Snodgrass1984, Schroeter1985, {Balthasar1986}}, and although newer technology enables high cadence observations, current studies have kept to the traditional cadences \citep[e.g.,][]{Kutsenko2022, Osipova2022}, seemingly without scientific justification. The effect of higher cadences on the accuracy of the retrieved rotation rate was explored in Sect.~4.2 of \citet{Emily_Thesis}, which showed that sunspot evolution and motions related to differential rotation cannot be separated for high-cadence time series. Thus, we settle on a slightly higher cadence of 12~hours, as this provides a better sampling close to the limb, while avoiding the aforementioned effects. 

For each sunspot, the rotation rate was determined by tracking its center over a given time after the spot was isolated from the rest of the disk. The process of finding the center of a sunspot starts with Gaussian smoothing to blur smaller features and granules. Next, a threshold was used to create a binary image. The threshold was set to 0.72, corresponding to the relative image intensity (with 1 representing the mean quiet-Sun level) after limb-darkening correction. This contrast threshold was applied uniformly across the disk. After eroding smaller features and finding the largest contour, the center coordinates of the binary sunspot area were determined using two methods. The first method determines the central moments of the binary image and calculates the center analog to a center of gravity, while the second method fits an ellipse of minimal area around the contours and uses the ellipse equation to determine the center. Both methods yield a pair of coordinates for the sunspot at each observed time, and both consistently showed similar results with an estimated error of $\pm 0.04^\circ$ per day respectively, demonstrating that they are equally reliable for determining sunspot positions and rotation rates. Therefore, the average coordinates of both methods were used for further evaluation. These coordinates were transformed into heliographical coordinates in the Carrington system \citep[see e.g.,][]{Balthasar1979, Thompson_2006}, which were used together with the timestamps of the individual images to determine the rotation rate during the spot's disk passage. The mean angular velocity of each sunspot, as well as the standard deviation of the mean, was saved alongside their assigned grade, which puts more statistical weight on sunspots of higher grade when fitting a rotation curve.

\subsection{Rotation curve fitting}\label{sec:2.2}

\begin{figure}[t!]
    \includegraphics[width=0.99\linewidth]{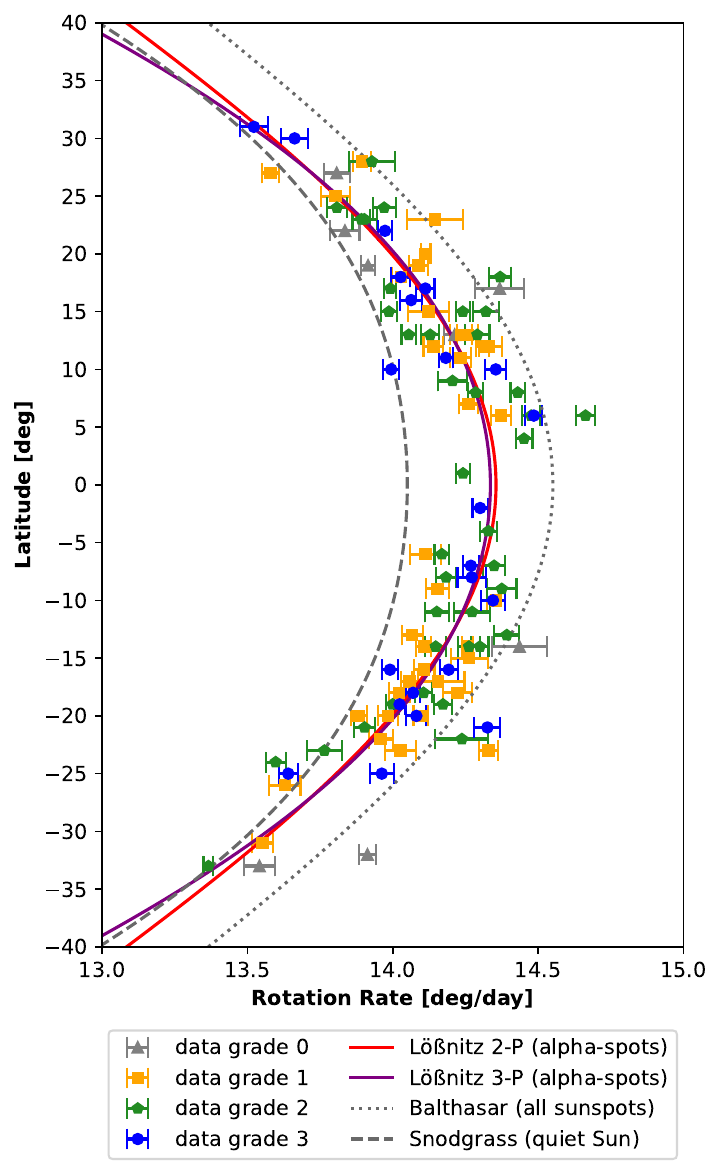}
    \caption{Rotation rate measurement across latitude of 105 sunspots between September 2013 and January 2024. Colors and symbols represent the grade of the sunspot (see Fig.~\ref{fig:sunspot_scores}). The measurements were fit according to Eq.~\ref{eq:1} by two rotation laws with two (red) or three (purple) parameters, respectively, and compared to other studies of all sunspot types (dotted) by \citet{Balthasar1986} and quiet Sun rotation via Doppler-measurements (dashed) by \citet{Snodgrass1984}.}
    \label{fig:fit_rotation_laws} 
\end{figure}

The rotation rates of individual sunspots are plotted in Fig.~\ref{fig:fit_rotation_laws} against their latitude and weighted by their respective sunspot grade. Sunspots, which turned out to represent simple, round $\alpha$ spots, were given a grade 0 and were excluded from fitting, while model spots of grade 3 have the highest influence on the resulting fit. From this velocity distribution, the rotation rate of the sample can be fit. The differential rotation law is commonly given in the form of
\begin{equation}
    \omega(\theta) = A + B \cdot \sin^2{\theta} + C \cdot \sin^4{\theta},
    \label{eq:1}
\end{equation}
where the coefficients $A$, $B$, and $C$ are determined by using a least-squares fit. The coefficient $A$ expresses the rotation rate at the equator and the coefficients $B$ and $C$ describe the differential shear toward higher latitudes. These parameters are given in Table~\ref{tab:SolarRoationCoefficients_COMPARISON}, along with other studies using the same standard rotation model (Eq.~\ref{eq:1}).

In rotation studies employing sunspots as photospheric tracers, the $C$ coefficient is often omitted, as it primarily affects the high-latitude behavior of the rotation law and the majority of sunspots appear within $\pm$35$^\circ$. Anything above that is exceedingly rare, to the point that \citet{Kejung2000} found that on average only 73 out of roughly 2000 groups per solar cycle form at a high latitude (above 35$^\circ$). The highest-latitude sunspot ever recorded had a latitude of 52$^\circ$ \citep{Dodson1953}.

This means that the use of a two-parameter law  for active regions on the Sun rests on the implicit assumption that this law's validity is restricted to the solar activity band. If a two-parameter law is used for the full range, it results in a less steep decline in velocity near the poles when compared to a three-parameter law that is more commonly used for the quiet Sun. For this work, we fit both a two- and three-parameter law to our sample of $\alpha$ sunspots and compare both forms with laws obtained by other studies, discussing the implications of our laws in the solar activity band, as well as their application and validity in higher latitude ranges (see Sect.~\ref{sec:Extrapolation}). For a discussion of the solar differential rotation of $\alpha$ sunspot in comparison to other spots, features, or the quiet Sun, we plot our rotation laws along side a law for all sunspot types by \citet{Balthasar1986} and quiet Sun rotation obtained from Doppler measurements by \citet{Snodgrass1984}.

\subsection{Extrapolation and scaling for other stars}
\label{sec:Extrapolation}

While spots are rarely observed beyond mid-latitudes on the Sun, on other stars high latitude and even polar spots have been reported \citep[e.g.,][]{Schrijver2001, Waite2011}. To enable meaningful comparisons between the solar and stellar spot rotation, we require a rotation law specifically tailored to sunspots (particularly $\alpha$ sunspots), rather than to other features or the quiet Sun. 

\begin{figure}[t!]
    \includegraphics[width=0.99\linewidth]{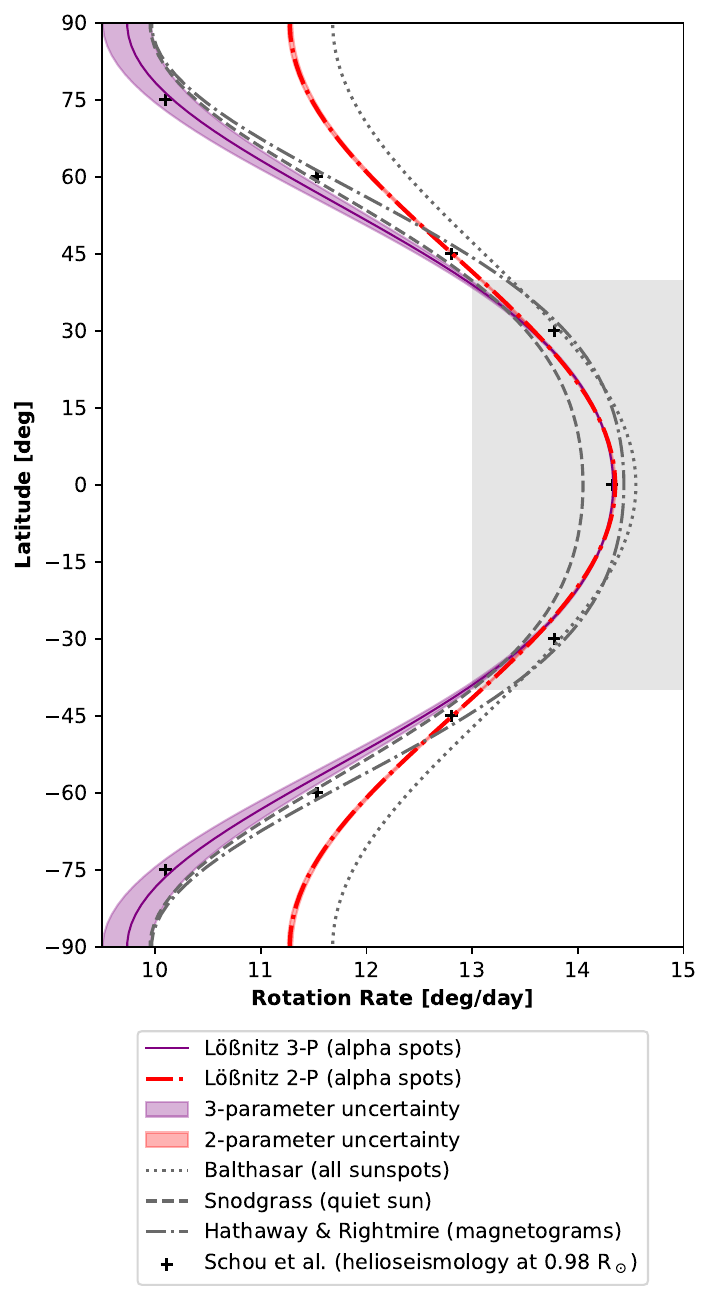}
    \caption{Our rotation laws with two (red) or three (purple) parameters for the full latitude range in comparison to a study of the rotation rate of the quiet Sun (dashed) by \citet{Snodgrass1984}, magnetic features and cell movement (dash-dotted) by \citet{Hathaway2011}, and a study of all sunspots classes (dotted) by \citet{Balthasar1986}. Additionally data derived from helioseismology studies of the internal solar rotation rates (pluses) by \cite{SchouHowe2002} has been added. The parameters of each mentioned study together with additional studies are given in Table~\ref{tab:SolarRoationCoefficients_COMPARISON}. The gray box shows the part of the plot previously shown in Fig. \ref{fig:fit_rotation_laws}. The uncertainties of the two- and three-parameter laws were estimated using a Monte-Carlo analysis. } 
    \label{fig:differential_rotation_plot}
\end{figure}

Although sunspots from our sample are not directly observed near the solar poles, we can still model and extrapolate their rotational behavior to higher latitudes using a fit of all three parameters. The lack of high-latitude data introduces higher uncertainties toward the poles, which we inferred via a Monte-Carlo analysis, whereby random perturbations of the input parameters were propagated through the model to generate a distribution of possible outcomes, defining the uncertainty range. Due to the constraints of the $A$ coefficient and the correlation with the $B$ coefficient, the uncertainty for the two-parameter law is expected to be very small in this case, while a three-parameter law will have a wider range closer to the poles. We were able to construct a rotation profile that could be reasonably compared to the behavior of polar spots on other stars. Other solar features, such as the large-scale convective cells studied by \citet{Hathaway2021}, also occur at high latitudes and extend to larger depths, exhibiting distinct rotation rates consistent with the deeper solar interior. To verify the validity of the extrapolation of our rotation law to high latitudes, we plot the full latitude range and compare the laws to the high-latitude behavior in other solar tracers, quiet-Sun rotation, and helioseismic measurements in Fig~\ref{fig:differential_rotation_plot} (see Sect.~\ref{sec:RotationLawResults}). 

For application to unresolved stellar observations, however, complex rotation profiles with multiple free parameters are poorly constrained and can introduce degeneracies.  This is likely the reason why most stellar differential rotation studies use a two-parameter rotation law, which introduces strong biases toward lower shear values on other stars, especially if they have high-latitude spots \citep[e.g.,][]{Petit2002, Reiners2007, Reinhold2013}.

Any fits should therefore be carried out with as few free parameters as possible while still accurately describing the physics. For this reason, we propose a modified two-parameter law whereby the $B$ and $C$ parameters are frozen to those of a solar law, and the curve itself is scaled with two new parameters,
\begin{equation}
    \omega(\theta) = \alpha + \beta ( B \sin^2{\theta} + C  \sin^4{\theta} ).\label{eq:stellar}
\end{equation}
Here, $\alpha$ is the rigid rotation rate of the star, and $\beta$ is the scaling factor for the differential rotation rate, which we call the corrected shear factor.

By changing the $\alpha$ parameter to below unity, subsolar differential rotators and even rigid rotators can be obtained. The opposite is true when $\alpha$ is above unity, creating super-solar rotators (see Fig.~\ref{fig:sage2}). This approach assumes that all stars have similar differential rotation curves despite having strongly differing rotation periods, which appears to be a valid assumption for G-type stars, but breaks down for hotter stars \citep{Reinhold2013}. Based on this assumption, we speculate that most solar-type and colder stars would have $\beta \approx 1$, while hotter stars and fast rotators would have a lower value. 

Depending on the type of study, we propose that the underlying solar law could be chosen to match the desired feature. For modeled stellar spots, we use our three-parameter $\alpha$-spot rotation law.

This proposed law is mathematically identical to the ``stiffened'' law used at Mt.\ Wilson Observatory to study changes in solar differential rotation over time \citep{Ulrich1988, Snodgrass1992}. However, in our definition the $B$ and $C$ coefficients are preserved and the $\beta$ coefficient is a dimensionless scalar. This law was one of the many attempts to separate the Sun's rotational pattern from other processes, and to resolve the cross-talk between the $A$, $B$, and $C$ parameters. For example, Gegenbauer polynomials were used to not only fit the  $A$, $B$, and $C$ parameters, but also meridional flows and instrumental effects, such as including stray light and instrumental drifts \citep[e.g.,][ Eq.~10]{Ulrich1988}. Versions of this approach are still used today \citep{Rao2024} in studies that have continuous latitude coverage.

\subsection{Implementation into the SAGE code}

To better illustrate the impact of differential rotation on stars, we implemented the scaling law from Eq.~\ref{eq:stellar} in the Stellar Activity Grid for Exoplanets \citep[SAGE, ][]{Hritam2024} code. SAGE uses a pixelation approach to project a three-dimensional stellar sphere on a two-dimensional Cartesian grid, along with the active regions and stellar atmospheric effects such as limb darkening and rotational broadening. We included the effect of differential rotation by adding the $\alpha$ and $\beta$ parameters. We created a toy model that consists of several stars with the same parameters and spot configurations, but different corrected shear factors. Subsequently, we obtained a model light curve for each star by rotating the stellar grid at the rotation period of the star.

\section{Results and discussion}

In this section, we discuss the rotation laws obtained from sunspot tracking and implement our proposed extrapolated law to the SAGE code and study its effect of differential rotation in light-curve modulation.

\subsection{Sunspot rotation law}
\label{sec:RotationLawResults}

The results of the sunspot tracking have been plotted in Fig.~\ref{fig:fit_rotation_laws}, showing each sunspot's average rotation rate plotted against its latitude. The color-coded $\alpha$-spot grades were included for better assessment, and although the sunspots with a grade of 0 were omitted during the fitting process, they are included for completeness. We find no significant trend in the rotation rates between our $\alpha$-spot classes. The parameters obtained from fitting both the two- and three-parameter laws to our sample are listed in Table~\ref{tab:SolarRoationCoefficients_COMPARISON} together with parameters obtained from other studies using Doppler- and tracer-based methods as well as helioseismology. We added four more rotation laws to the figure; namely, the curve of \citet{Balthasar1986} that represents the average rotation rate of sunspots of all types, a rotation law by \cite{Hathaway2011} that tracks magnetic elements, the quiet-Sun rotation curve by \citet{Snodgrass1984}, and a derived law from a helioseismic study by \citep[][Fig. 1]{SchouHowe2002}. 

Comparing rotation curves at the disk center with our fit results, we find that $\alpha$ sunspots are 1.35\% slower than average spots, but 1.56\% faster than the rotation of the quiet Sun. This corresponds to an average rooting depth at 0.98~$R_\odot$ or 13.9~Mm under the solar surface for $\alpha$ spots when matching the rate to the inner rotation rates reported by \citet{Howe2000} and \citet{SchouHowe2002}. The fact that our spots are slower than the general spot population is in line with the assumptions that they are older or that they are recurrent spots on average. This rooting depth is less deep than the value of 0.93~$R_\odot$ that was reported by \citet{Beck2000} and by \citet{Balthasar2007} for most (younger) sunspots. This further confirms that $\alpha$ spots are typically older and more detached compared to other spots. This trend is also consistent with other studies of sunspot (groups) and quiet photosphere measurements that are given in Table~\ref{tab:SolarRoationCoefficients_COMPARISON}. Our two-parameter law is similar to the two-parameter law of \citet{Balthasar1986} for Zürich-type H- and J-spots according to the classification by \citet{Waldmeier1955}. These two classes overlap with the \citet{Hale1919} type $\alpha$ spots and are therefore well comparable. 

However, as we look at higher latitudes in Fig. \ref{fig:differential_rotation_plot}, our three-parameter law declines more sharply than the quiet Sun law, resulting in sunspots that move more slowly than the surrounding quiet Sun at the poles. The resulting difference between the two- and three-parameter models increases significantly at higher latitudes, with the two-parameter law predicting rotation rates up to 2$^\circ$ per day faster. Comparing the curves with features such as giant granular cells or magnetic tracers from studies by \citet{Hathaway2011} and by \citet{Hathaway2021} shows that the two-parameter law would see spots moving around 1.5$^\circ$ per day (or $\approx$ 15\%) faster near the poles, whereas near the equator this difference to \citet{Hathaway2011} is only 0.5$^\circ$ per day (see Fig.~\ref{fig:differential_rotation_plot}). The three-parameter law, despite a systematic offset in rotation rate, follows the differential shear, which is consistent with that observed in magnetic features up to high latitudes \citep{Hathaway2011}. While the dynamics of giant cells or magnetic features are not directly comparable to those of sunspots, these findings support the idea that the structured differential rotation of features persist in polar regions and might therefore apply to polar sunspots on other stars.

A comparison to results from helioseismology further justifies the extrapolated three-parameter law, as can be seen with the data in Fig. \ref{fig:differential_rotation_plot}. At a depth of 0.98~$R_\odot$, results from \citet{SchouHowe2002} show the same trend as our three-parameter law at high latitudes. Due to the strong connection between ($\alpha$-)sunspots and the rotation rates of subsurface layers, this agreement to our model further supports the usage and physical validity of a three-parameter law based on sunspots as a basis for stellar spot rotation, while a two-parameter law is insufficient to model stellar spots at higher latitudes. We therefore use the three-parameter law in our stellar toy model.

\begin{figure}[t!]
    \includegraphics[width=0.99\columnwidth]{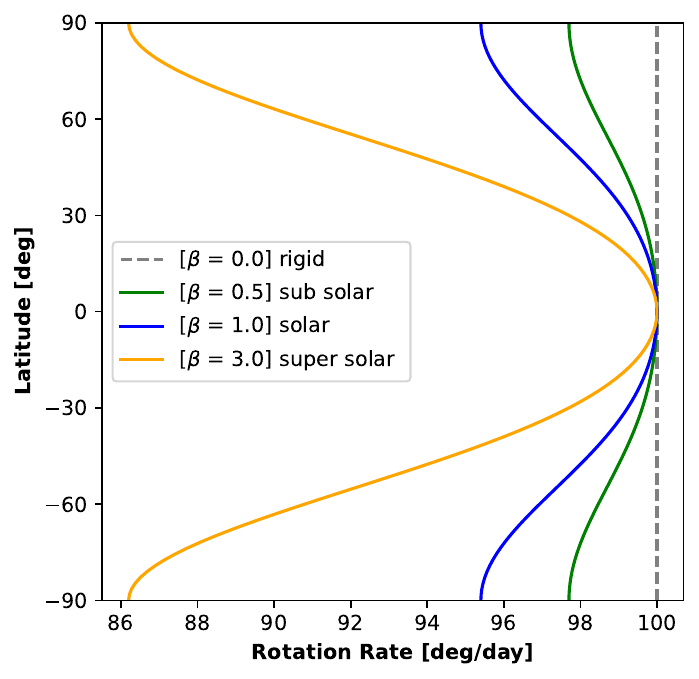}
    \caption{Scaled differential rotation curves for a two-day period rotator according to Eq.~\ref{eq:sclaing_law}, showing the different rotation profiles for a super (orange), subsolar (green), and solar (blue) rotator, as well as a rigid rotation curve (dashed).}
    \label{fig:sage2}
\end{figure}

\begin{figure*}[t!]
    \includegraphics[width=0.96\textwidth, trim={5.0cm 0.0cm 4.5cm 0.0cm},clip]{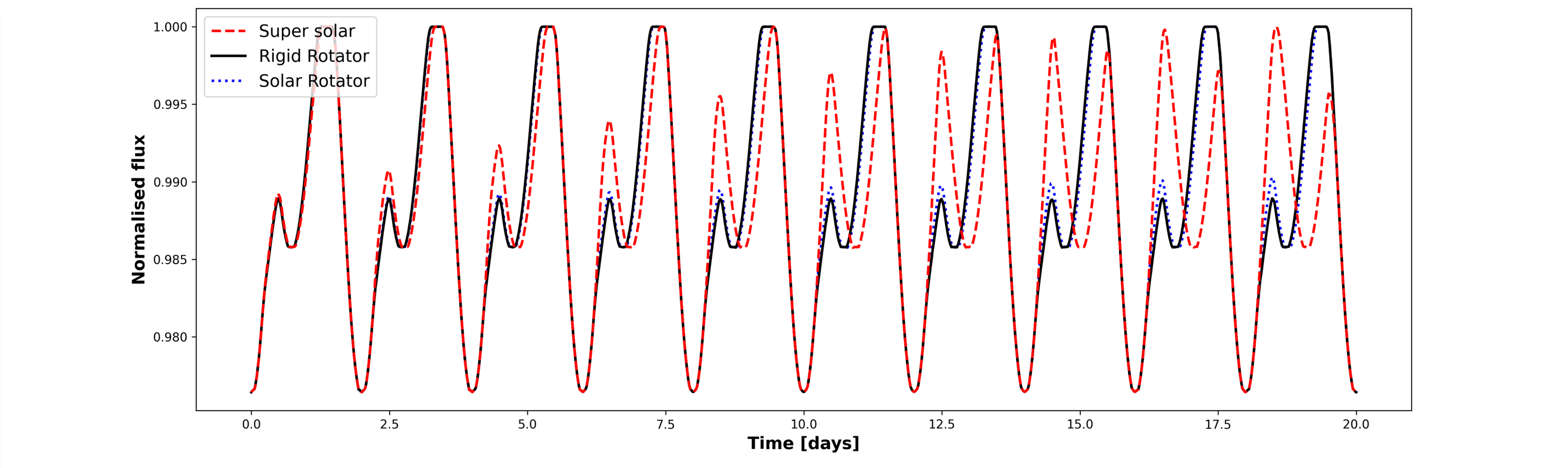}
    \caption{Light curves produced by a rigid rotator (solid black curve), a super solar rotator (dashed red curve), and a solar rotator (dotted blue curve) over the course of ten rotations.}
    \label{fig:sage1}
\end{figure*}

\subsection{Differential rotation on other stars}

As is discussed in Sect.~\ref{sec:Extrapolation}, we transformed our three-parameter rotation law into  
\begin{align}
    \omega(\theta) = \alpha - \beta ( 2.56 \sin^2{\theta} + 2.04  \sin^4{\theta} ),
    \label{eq:sclaing_law}
\end{align}
using the values from our three-parameter rotation law.

To illustrate the effect of differential rotation on light-curve modulation, we created a stellar grid for a G2V-type star with two spots at different latitudes (0$^{\circ}$ and 30$^{\circ}$) and assuming a spot contrast of 0.7. Two spots with angular sizes of 6$^{\circ}$ and 12$^{\circ}$, respectively, were modeled. A set of four scaled laws was created with a corrected shear factor, $\beta$, of 0.0, 0.5, 1.0, and 3.0 for rigid rotators, subsolar rotators, solar rotators, and super-solar rotators, respectively (see Fig.~\ref{fig:sage2}).

In Fig.~\ref{fig:sage1}, we see ten rotations of the star, which starts with one spot on the front and a smaller one on the back at a higher latitude. The relative distance between the spots changes as a result of the differential rotation laws. We see a rigid rotator in black, a super-solar rotator in dashed red, and a solar rotator in dotted blue. A small difference can be seen in the blue curve with respect to the rigid rotator and a much larger change in the red curve. This modulation in the light-curve signal is reminiscent of so-called ``scallop-shell'' light curves \citep{Stauffer2017} whereby the light curve changes slowly over time. While such light curves are typically explained with clouds and dense prominences, we propose that differential rotation may also cause similar patterns.

\section{Conclusions}\label{conclusions}

In this study, we have presented a detailed investigation of the rotation rates of $\alpha$ sunspots using high-cadence data from SDO/HMI, with a particular focus on their role as tracers of solar differential rotation. Our manual selection and tracking of 105 well-defined $\alpha$ spots allowed for robust angular velocity measurements with the {\tt SpotiPy} code. The differential rotation law as a function of the heliographic latitude for these spots was determined to be 
\begin{align}
    \omega (\theta) = ( 14.336 -2.561 \sin^2{\theta} - 2.04 \sin^4{\theta})\, \mathrm{deg~day}^{-1}
\end{align}
with a three-parameter model and
\begin{align}
    \omega (\theta) = ( 14.356 -3.07 \sin^2{\theta})\, \mathrm{deg~day}^{-1}\,
\end{align}
with the two-parameter model. The latter is more commonly used regarding sunspot-based rotation curves in solar physics, as it accurately describes the spots on the solar activity belt with fewer variables, but cannot be considered valid outside of this range, as it does not describe the differential rotation profiles of any other solar feature in high latitudes. It can therefore not be generally applied to starspots.

These results show that $\alpha$ spots exhibit slightly lower rotation rates when compared to other sunspot populations but still rotate faster than the surrounding quiet photosphere. This is in line with the idea that older spots tend to be rooted less deeply than younger ones. For our sample, we find that on average they are rooted at 0.98~$R_\odot$ or 13.9~Mm below the solar surface. We therefore caution that stellar rotation rates inferred from spot tracking may be systematically overestimated, as stellar spots -- such as sunspots -- can rotate faster than the stellar surface due to their connection to deeper, more rapidly rotating layers.

Building on these findings, we used the differential rotation law based specifically on $\alpha$ spots, contributing a tracer-specific solar rotation profile that may prove useful for cross-comparisons with other solar and stellar datasets. Extending the solar analysis to a stellar context, we propose a physically motivated scaling approach that links solar differential rotation characteristics to stars of varying rotational regimes, which scales our three-parameter law with two parameters to recreate sub- and super-solar rotation curves (see Fig.~\ref{fig:sage2} and Eq.~\ref{eq:stellar}). We implemented this law in the SAGE simulation framework to evaluate the effects of differential rotation on stellar light curves, showing that it can significantly influence photometric variability and rotation period estimates and that it cause patterns similar to scallop-shell light curves, which are traditionally explained with prominences.

This work combines detailed solar observations with stellar modeling efforts, highlighting the value of $\alpha$ spots as a stable and interpretable rotation tracer. The combination of observational analysis and forward modeling provides a base for improving our understanding of surface differential rotation in stars and its influence on exoplanet transit measurements. We plan to follow up this work by applying our scaled differential rotation law to stellar light curves with multiple spots.

\begin{acknowledgements}
This work is based on a bachelor's thesis by EJL \citep{Emily_Thesis}.
We thank Dr. Charles Baldner for a stimulating discussion on the creation of HMI limb-darkening free images. We thank Dr. Jake Mitchell for helpful comments and suggestions on the manuscript.
AP is supported by the \emph{Deut\-sche For\-schungs\-ge\-mein\-schaft, DFG\/} project number PI 2102/1-1

This research has made use of NASA's Astrophysics Data System (ADS) bibliographic services. 
We acknowledge the community efforts devoted to the development of the following open-source packages that were used in this work: numpy (\href{http:\\numpy.org}{numpy.org}), matplotlib (\href{http:\\matplotlib.org}{matplotlib.org}), and astropy (\href{http:\\astropy.org}{astropy.org}).
We extensively used SunPy \citep{Sunpy}, the ISPy library \citep{ISPy2021}, and SOAImage DS9 \citep{2003DS9} for data visualization and processing.

\end{acknowledgements}

\bibliographystyle{aa}
\bibliography{ref}

@article{Kejung2000,
    author = {Kejun, Li and Xiaoma, Gu and Fuyuan, Xiang and Xiaohua, Liu and Xuekun, Chen},
    title = {Sunspot groups at high latitude},
    journal = {Monthly Notices of the Royal Astronomical Society},
    volume = {317},
    number = {4},
    pages = {897-901},
    year = {2000},
    month = {10},
    issn = {0035-8711},
    doi = {10.1046/j.1365-8711.2000.03758.x},
    url = {https://doi.org/10.1046/j.1365-8711.2000.03758.x},
    eprint = {https://academic.oup.com/mnras/article-pdf/317/4/897/3499665/317-4-897.pdf},
}

@ARTICLE{Dodson1953,
       author = {{Dodson}, Helen W.},
        title = "{High Latitude Sunspot, August 13, 1953}",
      journal = {\pasp},
         year = 1953,
        month = oct,
       volume = {65},
       number = {386},
        pages = {256},
          doi = {10.1086/126615},
       adsurl = {https://ui.adsabs.harvard.edu/abs/1953PASP...65..256D}
}

@article{Schmassmann2021,
	author = {{Schmassmann}, M. and {Rempel}, M. and {Bello González}, N. and {Schlichenmaier}, R. and {Jurčák}, J.},
	title = {Characterization of magneto-convection in sunspots - The Gough-Tayler stability criterion in MURaM sunspot simulations⋆},
	DOI= "10.1051/0004-6361/202141607",
	url= "https://doi.org/10.1051/0004-6361/202141607",
	journal = {A\&A},
	year = 2021,
	volume = 656,
	pages = "A92",
}

@ARTICLE{Rempel2009,
       author = {{Rempel}, M. and {Sch{\"u}ssler}, M. and {Kn{\"o}lker}, M.},
        title = "{Radiative Magnetohydrodynamic Simulation of Sunspot Structure}",
      journal = {\apj},
         year = 2009,
        month = jan,
       volume = {691},
       number = {1},
        pages = {640-649},
          doi = {10.1088/0004-637X/691/1/640},
archivePrefix = {arXiv},
       eprint = {0808.3294},
 primaryClass = {astro-ph},
       adsurl = {https://ui.adsabs.harvard.edu/abs/2009ApJ...691..640R}
}

@ARTICLE{SchouHowe2002,
       author = {{Schou}, J. and {Howe}, R. and {Basu}, S. and {Christensen-Dalsgaard}, J. and {Corbard}, T. and {Hill}, F. and {Komm}, R. and {Larsen}, R.~M. and {Rabello-Soares}, M.~C. and {Thompson}, M.~J.},
        title = "{A Comparison of Solar p-Mode Parameters from the Michelson Doppler Imager and the Global Oscillation Network Group: Splitting Coefficients and Rotation Inversions}",
      journal = {\apj},
         year = 2002,
        month = mar,
       volume = {567},
       number = {2},
        pages = {1234-1249},
          doi = {10.1086/338665},
       adsurl = {https://ui.adsabs.harvard.edu/abs/2002ApJ...567.1234S}
}

@ARTICLE{Barnes2016,
       author = {{Barnes}, Sydney A. and {Weingrill}, Joerg and {Fritzewski}, Dario and {Strassmeier}, Klaus G. and {Platais}, Imants},
        title = "{Rotation Periods for Cool Stars in the 4 Gyr old Open Cluster M67, The Solar-Stellar Connection, and the Applicability of Gyrochronology to at least Solar Age}",
      journal = {\apj},
         year = 2016,
        month = may,
       volume = {823},
       number = {1},
          eid = {16},
        pages = {16},
          doi = {10.3847/0004-637X/823/1/16},
archivePrefix = {arXiv},
       eprint = {1603.09179},
 primaryClass = {astro-ph.SR},
       adsurl = {https://ui.adsabs.harvard.edu/abs/2016ApJ...823...16B}
}

@ARTICLE{Ulrich1988,
       author = {{Ulrich}, Roger K. and {Boyden}, John E. and {Webster}, Larry and {Snodgrass}, Herschel B. and {Padilla}, Steven P. and {Gilman}, Pamela and {Shieber}, Tom},
        title = "{Solar Rotation Measurements at MT.WILSON - Part Five}",
      journal = {\solphys},
         year = 1988,
        month = sep,
       volume = {117},
       number = {2},
        pages = {291-328},
          doi = {10.1007/BF00147250},
       adsurl = {https://ui.adsabs.harvard.edu/abs/1988SoPh..117..291U}
}

@INPROCEEDINGS{Snodgrass1992,
       author = {{Snodgrass}, Herschel B.},
        title = "{Synoptic Observations of Large Scale Velocity Patterns on the Sun}",
    booktitle = {The solar cycle},
         year = 1992,
       editor = {{Harvey}, Karen L.},
       series = {ASP Conf. Ser.},
       volume = {27},
        month = jan,
        pages = {205},
       adsurl = {https://ui.adsabs.harvard.edu/abs/1992ASPC...27..205S}
}

@article{Beck2000,
author={Beck, John G.},
title={A comparison of differential rotation measurements -- (Invited Review)},
journal={\solphys},
year={2000},
month={Jan},
day={01},
volume={191},
number={1},
pages={47-70},
issn={1573-093X},
doi={10.1023/A:1005226402796},
url={https://doi.org/10.1023/A:1005226402796}
}

@article{Ward1966,
author={Ward, Fred},
title={The longitudinal proper motion of sunspots and the solar rotation rate},
journal="{Pure Appl. Geophys.}",
year={1966},
month={Dec},
day={01},
volume={63},
number={1},
pages={196-204},
issn={1420-9136},
doi={10.1007/BF00875169},
url={https://doi.org/10.1007/BF00875169}
}

@article{Tlatov2024,
doi = {10.3847/1538-4357/ad90a0},
url = {https://dx.doi.org/10.3847/1538-4357/ad90a0},
year = {2024},
month = {dec},
publisher = {The American Astronomical Society},
volume = {977},
number = {1},
pages = {110},
author = {Tlatov, Andrey G. and Tlatova, Kseniya A.},
title = {Differential Rotation of Individual Sunspots and Pores},
journal = {\apj}
}

@article{Howe2000,
author = {R. Howe  and J. Christensen-Dalsgaard  and F. Hill  and R. W. Komm  and R. M. Larsen  and J. Schou  and M. J. Thompson  and J. Toomre },
title = {Dynamic Variations at the Base of the Solar Convection Zone},
journal = {Science},
volume = {287},
number = {5462},
pages = {2456-2460},
year = {2000},
doi = {10.1126/science.287.5462.2456},
URL = {https://www.science.org/doi/abs/10.1126/science.287.5462.2456},
eprint = {https://www.science.org/doi/pdf/10.1126/science.287.5462.2456}}

@ARTICLE{Sunpy,
  doi = {10.3847/1538-4357/ab4f7a},
  url = {https://iopscience.iop.org/article/10.3847/1538-4357/ab4f7a},
  author = {{The SunPy Community} and Barnes, Will T. and Bobra, Monica G. and Christe, Steven D. and Freij, Nabil and Hayes, Laura A. and Ireland, Jack and Mumford, Stuart and Perez-Suarez, David and Ryan, Daniel F. and Shih, Albert Y. and Chanda, Prateek and Glogowski, Kolja and Hewett, Russell and Hughitt, V. Keith and Hill, Andrew and Hiware, Kaustubh and Inglis, Andrew and Kirk, Michael S. F. and Konge, Sudarshan and Mason, James Paul and Maloney, Shane Anthony and Murray, Sophie A. and Panda, Asish and Park, Jongyeob and Pereira, Tiago M. D. and Reardon, Kevin and Savage, Sabrina and Sipőcz, Brigitta M. and Stansby, David and Jain, Yash and Taylor, Garrison and Yadav, Tannmay and Rajul and Dang, Trung Kien},
  title = {The SunPy Project: Open Source Development and Status of the Version 1.0 Core Package},
  journal = {\apj},
  volume = {890},
  issue = {1},
  pages = {68-},
  publisher = {American Astronomical Society},
  year = {2020}
}

@ARTICLE{Donati1997,
       author = {{Donati}, J. -F. and {Collier Cameron}, A.},
        title = "{Differential rotation and magnetic polarity patterns on AB Doradus}",
      journal = {\mnras},
         year = 1997,
        month = oct,
       volume = {291},
       number = {1},
        pages = {1-19},
          doi = {10.1093/mnras/291.1.1},
       adsurl = {https://ui.adsabs.harvard.edu/abs/1997MNRAS.291....1D}
}

@ARTICLE{Zaleski2020,
       author = {{Zaleski}, S.~M. and {Valio}, A. and {Carter}, B.~D. and {Marsden}, S.~C.},
        title = "{Activity and differential rotation of the early M dwarf Kepler-45 from transit mapping}",
      journal = {\mnras},
         year = 2020,
        month = mar,
       volume = {492},
       number = {4},
        pages = {5141-5151},
          doi = {10.1093/mnras/staa103},
       adsurl = {https://ui.adsabs.harvard.edu/abs/2020MNRAS.492.5141Z}
}

@ARTICLE{Cauley2018,
       author = {{Cauley}, P. Wilson and {Kuckein}, Christoph and {Redfield}, Seth and {Shkolnik}, Evgenya L. and {Denker}, Carsten and {Llama}, Joe and {Verma}, Meetu},
        title = "{The Effects of Stellar Activity on Optical High-resolution Exoplanet Transmission Spectra}",
      journal = {\aj},
         year = 2018,
        month = nov,
       volume = {156},
       number = {5},
          eid = {189},
        pages = {189},
          doi = {10.3847/1538-3881/aaddf9},
archivePrefix = {arXiv},
       eprint = {1808.09558},
 primaryClass = {astro-ph.EP},
       adsurl = {https://ui.adsabs.harvard.edu/abs/2018AJ....156..189C}
}

@ARTICLE{Petit2024,
       author = {{Petit dit de la Roche}, D.~J.~M. and {Chakraborty}, H. and {Lendl}, M. and {Kitzmann}, D. and {Pietrow}, A.~G.~M. and {Akinsanmi}, B. and {Boffin}, H.~M.~J. and {Cubillos}, Patricio E. and {Deline}, A. and {Ehrenreich}, D. and {Fossati}, L. and {Sedaghati}, E.},
        title = "{Detection of faculae in the transit and transmission spectrum of WASP-69b}",
      journal = {\aap},
         year = 2024,
        month = dec,
       volume = {692},
          eid = {A83},
        pages = {A83},
          doi = {10.1051/0004-6361/202451740},
archivePrefix = {arXiv},
       eprint = {2410.18663},
 primaryClass = {astro-ph.EP},
       adsurl = {https://ui.adsabs.harvard.edu/abs/2024A&A...692A..83P}
}

@ARTICLE{Soap2012,
       author = {{Boisse}, I. and {Bonfils}, X. and {Santos}, N.~C.},
        title = "{SOAP. A tool for the fast computation of photometry and radial velocity induced by stellar spots}",
      journal = {\aap},
         year = 2012,
        month = sep,
       volume = {545},
          eid = {A109},
        pages = {A109},
          doi = {10.1051/0004-6361/201219115},
archivePrefix = {arXiv},
       eprint = {1206.5493},
 primaryClass = {astro-ph.IM},
       adsurl = {https://ui.adsabs.harvard.edu/abs/2012A&A...545A.109B}
}

@misc{Emily_Thesis,
author = {Lößnitz, Emily Joe},
year = {2024},
month = {08},
school = {University Potsdam, Germany},
title = "{A study of solar differential rotation and limb-darkening in $\alpha$-sunspots}",
doi = {10.13140/RG.2.2.28680.33286},
note = "{Bachelor's thesis, at University Potsdam, Germany}"
}

@ARTICLE{Stauffer2017,
       author = {{Stauffer}, John and {Collier Cameron}, Andrew and {Jardine}, Moira and {David}, Trevor J. and {Rebull}, Luisa and {Cody}, Ann Marie and {Hillenbrand}, Lynne A. and {Barrado}, David and {Wolk}, Scott and {Davenport}, James and {Pinsonneault}, Marc},
        title = "{Orbiting Clouds of Material at the Keplerian Co-rotation Radius of Rapidly Rotating Low-mass WTTs in Upper Sco}",
      journal = {\aj},
         year = 2017,
        month = apr,
       volume = {153},
       number = {4},
          eid = {152},
        pages = {152},
          doi = {10.3847/1538-3881/aa5eb9},
archivePrefix = {arXiv},
       eprint = {1702.01797},
 primaryClass = {astro-ph.SR},
       adsurl = {https://ui.adsabs.harvard.edu/abs/2017AJ....153..152S}
}

@ARTICLE{Korhonen2021,
       author = {{Korhonen}, H. and {Roettenbacher}, R.~M. and {Gu}, S. and {Grundahl}, F. and {Andersen}, M.~F. and {Henry}, G.~W. and {Jessen-Hansen}, J. and {Antoci}, V. and {Pall{\'e}}, P.~L.},
        title = "{Observing the changing surface structures of the active K giant {\ensuremath{\sigma}} Geminorum with SONG}",
      journal = {\aap},
         year = 2021,
        month = feb,
       volume = {646},
          eid = {A6},
        pages = {A6},
          doi = {10.1051/0004-6361/202038799},
archivePrefix = {arXiv},
       eprint = {2012.15177},
 primaryClass = {astro-ph.SR},
       adsurl = {https://ui.adsabs.harvard.edu/abs/2021A&A...646A...6K}
}

@Article{Rao2024,
    author={Rao, Shihao
    and Li, Chuan
    and Ding, Mingde
    and Hong, Jie
    and Chen, Feng
    and Fang, Cheng
    and Qiu, Ye
    and Li, Zhen
    and Chen, Pengfei
    and Li, Kejun
    and Hao, Qi
    and Guo, Yang
    and Cheng, Xin
    and Dai, Yu
    and Peng, Zhixin
    and You, Wei
    and Yuan, Yuan},
    title={Height-dependent differential rotation of the solar atmosphere detected by CHASE},
    journal={Nat. Astron.},
    year={2024},
    month={Sep},
    day={01},
    volume={8},
    number={9},
    pages={1102-1109},
    issn={2397-3366},
    doi={10.1038/s41550-024-02299-4},
    url={https://doi.org/10.1038/s41550-024-02299-4}
}

@article{Li2024,
    author = {Li, K J and Xu, J C},
    title = {The differential rotation of the chromosphere and the quiet chromosphere in the falling and rising periods of a solar cycle},
    journal = {\mnras},
    volume = {528},
    number = {2},
    pages = {1438-1444},
    year = {2024},
    month = {01},
    issn = {0035-8711},
    doi = {10.1093/mnras/stae044},
    url = {https://doi.org/10.1093/mnras/stae044},
    eprint = {https://academic.oup.com/mnras/article-pdf/528/2/1438/56409461/stae044.pdf},
}

@article{Mishra2024,
doi = {10.3847/1538-4357/ad1188},
url = {https://dx.doi.org/10.3847/1538-4357/ad1188},
year = {2024},
month = {jan},
publisher = {The American Astronomical Society},
volume = {961},
number = {1},
pages = {40},
author = {Mishra, Dibya Kirti and Routh, Srinjana and Jha, Bibhuti Kumar and Chatzistergos, Theodosios and Basu, Judhajeet and Chatterjee, Subhamoy and Banerjee, Dipankar and Ermolli, Ilaria},
title = {Differential Rotation of the Solar Chromosphere: A Century-long Perspective from Kodaikanal Solar Observatory Ca ii K Data},
journal = {\apj},
}

@article{Wan2023,
    author = {Wan, M and Gao, P X and Xu, J C and Shi, X J and Xiang, N B and Xie, J L},
    title = {Differential rotation: the chromosphere to the quiet chromosphere},
    journal = {\mnras},
    volume = {520},
    number = {1},
    pages = {988-993},
    year = {2023},
    month = {01},
    issn = {0035-8711},
    doi = {10.1093/mnras/stad192},
    url = {https://doi.org/10.1093/mnras/stad192},
    eprint = {https://academic.oup.com/mnras/article-pdf/520/1/988/49058086/stad192.pdf},
}

@ARTICLE{Hathaway2011,
       author = {{Hathaway}, David H. and {Rightmire}, Lisa},
        title = "{Variations in the Axisymmetric Transport of Magnetic Elements on the Sun: 1996-2010}",
      journal = {\apj},
         year = 2011,
        month = mar,
       volume = {729},
       number = {2},
          eid = {80},
        pages = {80},
          doi = {10.1088/0004-637X/729/2/80},
archivePrefix = {arXiv},
       eprint = {1010.1242},
 primaryClass = {astro-ph.SR},
       adsurl = {https://ui.adsabs.harvard.edu/abs/2011ApJ...729...80H}
}

@ARTICLE{Hathaway2021,
       author = {{Hathaway}, David H. and {Upton}, Lisa A.},
        title = "{Hydrodynamic Properties of the Sun's Giant Cellular Flows}",
      journal = {\apj},
         year = 2021,
        month = feb,
       volume = {908},
       number = {2},
          eid = {160},
        pages = {160},
          doi = {10.3847/1538-4357/abcbfa},
archivePrefix = {arXiv},
       eprint = {2006.06084},
 primaryClass = {astro-ph.SR},
       adsurl = {https://ui.adsabs.harvard.edu/abs/2021ApJ...908..160H}
}

@INPROCEEDINGS{Cheops2013,
       author = {{Broeg}, C. and {Fortier}, A. and {Ehrenreich}, D. and {Alibert}, Y. and {Baumjohann}, W. and {Benz}, W. and {Deleuil}, M. and {Gillon}, M. and {Ivanov}, A. and {Liseau}, R. and {Meyer}, M. and {Oloffson}, G. and {Pagano}, I. and {Piotto}, G. and {Pollacco}, D. and {Queloz}, D. and {Ragazzoni}, R. and {Renotte}, E. and {Steller}, M. and {Thomas}, N.},
        title = "{CHEOPS: A transit photometry mission for ESA's small mission programme}",
    booktitle = {Hot Planets and Cool Stars},
         year = 2013,
       series = {Eur. Phys. J. Web Conf.},
       volume = {47},
        month = apr,
          eid = {03005},
        pages = {03005},
          doi = {10.1051/epjconf/20134703005},
archivePrefix = {arXiv},
       eprint = {1305.2270},
 primaryClass = {astro-ph.EP},
       adsurl = {https://ui.adsabs.harvard.edu/abs/2013EPJWC..4703005B}
}

@ARTICLE{Tess2015,
       author = {{Ricker}, George R. and {Winn}, Joshua N. and {Vanderspek}, Roland and {Latham}, David W. and {Bakos}, G{\'a}sp{\'a}r {\'A}. and {Bean}, Jacob L. and {Berta-Thompson}, Zachory K. and {Brown}, Timothy M. and {Buchhave}, Lars and {Butler}, Nathaniel R. and {Butler}, R. Paul and {Chaplin}, William J. and {Charbonneau}, David and {Christensen-Dalsgaard}, J{\o}rgen and {Clampin}, Mark and {Deming}, Drake and {Doty}, John and {De Lee}, Nathan and {Dressing}, Courtney and {Dunham}, Edward W. and {Endl}, Michael and {Fressin}, Francois and {Ge}, Jian and {Henning}, Thomas and {Holman}, Matthew J. and {Howard}, Andrew W. and {Ida}, Shigeru and {Jenkins}, Jon M. and {Jernigan}, Garrett and {Johnson}, John Asher and {Kaltenegger}, Lisa and {Kawai}, Nobuyuki and {Kjeldsen}, Hans and {Laughlin}, Gregory and {Levine}, Alan M. and {Lin}, Douglas and {Lissauer}, Jack J. and {MacQueen}, Phillip and {Marcy}, Geoffrey and {McCullough}, Peter R. and {Morton}, Timothy D. and {Narita}, Norio and {Paegert}, Martin and {Palle}, Enric and {Pepe}, Francesco and {Pepper}, Joshua and {Quirrenbach}, Andreas and {Rinehart}, Stephen A. and {Sasselov}, Dimitar and {Sato}, Bun'ei and {Seager}, Sara and {Sozzetti}, Alessandro and {Stassun}, Keivan G. and {Sullivan}, Peter and {Szentgyorgyi}, Andrew and {Torres}, Guillermo and {Udry}, Stephane and {Villasenor}, Joel},
        title = "{Transiting Exoplanet Survey Satellite (TESS)}",
      journal = {JATIS},
         year = 2015,
        month = jan,
       volume = {1},
          eid = {014003},
        pages = {014003},
          doi = {10.1117/1.JATIS.1.1.014003},
       adsurl = {https://ui.adsabs.harvard.edu/abs/2015JATIS...1a4003R}
}

@ARTICLE{2010Kepler,
       author = {{Borucki}, William J. and {Koch}, David and {Basri}, Gibor and {Batalha}, Natalie and {Brown}, Timothy and {Caldwell}, Douglas and {Caldwell}, John and {Christensen-Dalsgaard}, J{\o}rgen and {Cochran}, William D. and {DeVore}, Edna and {Dunham}, Edward W. and {Dupree}, Andrea K. and {Gautier}, Thomas N. and {Geary}, John C. and {Gilliland}, Ronald and {Gould}, Alan and {Howell}, Steve B. and {Jenkins}, Jon M. and {Kondo}, Yoji and {Latham}, David W. and {Marcy}, Geoffrey W. and {Meibom}, S{\o}ren and {Kjeldsen}, Hans and {Lissauer}, Jack J. and {Monet}, David G. and {Morrison}, David and {Sasselov}, Dimitar and {Tarter}, Jill and {Boss}, Alan and {Brownlee}, Don and {Owen}, Toby and {Buzasi}, Derek and {Charbonneau}, David and {Doyle}, Laurance and {Fortney}, Jonathan and {Ford}, Eric B. and {Holman}, Matthew J. and {Seager}, Sara and {Steffen}, Jason H. and {Welsh}, William F. and {Rowe}, Jason and {Anderson}, Howard and {Buchhave}, Lars and {Ciardi}, David and {Walkowicz}, Lucianne and {Sherry}, William and {Horch}, Elliott and {Isaacson}, Howard and {Everett}, Mark E. and {Fischer}, Debra and {Torres}, Guillermo and {Johnson}, John Asher and {Endl}, Michael and {MacQueen}, Phillip and {Bryson}, Stephen T. and {Dotson}, Jessie and {Haas}, Michael and {Kolodziejczak}, Jeffrey and {Van Cleve}, Jeffrey and {Chandrasekaran}, Hema and {Twicken}, Joseph D. and {Quintana}, Elisa V. and {Clarke}, Bruce D. and {Allen}, Christopher and {Li}, Jie and {Wu}, Haley and {Tenenbaum}, Peter and {Verner}, Ekaterina and {Bruhweiler}, Frederick and {Barnes}, Jason and {Prsa}, Andrej},
        title = "{Kepler Planet-Detection Mission: Introduction and First Results}",
      journal = {Science},
         year = 2010,
        month = feb,
       volume = {327},
       number = {5968},
        pages = {977},
          doi = {10.1126/science.1185402},
       adsurl = {https://ui.adsabs.harvard.edu/abs/2010Sci...327..977B}
}

@ARTICLE{Goldreich1967,
       author = {{Goldreich}, Peter and {Schubert}, Gerald},
        title = "{Differential Rotation in Stars}",
      journal = {\apj},
         year = 1967,
        month = nov,
       volume = {150},
        pages = {571},
          doi = {10.1086/149360},
       adsurl = {https://ui.adsabs.harvard.edu/abs/1967ApJ...150..571G}
}

@ARTICLE{Kutsenko2022,
       author = {{Kutsenko}, Alexander S. and {Abramenko}, Valentina I.},
        title = "{Probing the rotation rate of solar active regions: the comparison of methods}",
      journal = {Open Astron.},
         year = 2022,
        month = jan,
       volume = {30},
       number = {1},
        pages = {219-224},
          doi = {10.1515/astro-2021-0029},
       adsurl = {https://ui.adsabs.harvard.edu/abs/2022OAst...30..219K}
}

@ARTICLE{Reiners2002,
       author = {{Reiners}, A. and {Schmitt}, J.~H.~M.~M.},
        title = "{On the feasibility of the detection of differential rotation in stellar absorption profiles}",
      journal = {\aap},
         year = 2002,
        month = mar,
       volume = {384},
        pages = {155-162},
          doi = {10.1051/0004-6361:20011801},
       adsurl = {https://ui.adsabs.harvard.edu/abs/2002A&A...384..155R}
}

@ARTICLE{Balthasar1986,
       author = {{Balthasar}, H. and {Vazquez}, M. and {Wöhl}, H.},
        title = "{Differential rotation of sunspot groups in the period from 1874 through 1976 and changes of the rotation velocity within the solar cycle}",
      journal = {\aap},
         year = 1986,
        month = jan,
       volume = {155},
       number = {1},
        pages = {87-98},
       adsurl = {https://ui.adsabs.harvard.edu/abs/1986A&A...155...87B}
}

@ARTICLE{Balthasar1982,
       author = {{Balthasar}, H. and {Schüssler}, M. and {Wöhl}, H.},
        title = "{On changes of the rotation velocities of stable, recurrent sunspots and their interpretation with a flux tube model}",
      journal = {\solphys},
         year = 1982,
        month = feb,
       volume = {76},
       number = {1},
        pages = {21-28},
          doi = {10.1007/BF00214127},
       adsurl = {https://ui.adsabs.harvard.edu/abs/1982SoPh...76...21B}
}

@ARTICLE{Netto2020,
       author = {{Netto}, Y. and {Valio}, A.},
        title = "{Stellar magnetic activity and the butterfly diagram of Kepler-63}",
      journal = {\aap},
         year = 2020,
        month = mar,
       volume = {635},
          eid = {A78},
        pages = {A78},
          doi = {10.1051/0004-6361/201936219},
archivePrefix = {arXiv},
       eprint = {1911.08661},
 primaryClass = {astro-ph.SR},
       adsurl = {https://ui.adsabs.harvard.edu/abs/2020A&A...635A..78N}
}

@ARTICLE{Reinhold2013,
       author = {{Reinhold}, Timo and {Reiners}, Ansgar and {Basri}, Gibor},
        title = "{Rotation and differential rotation of active Kepler stars}",
      journal = {\aap},
         year = 2013,
        month = dec,
       volume = {560},
          eid = {A4},
        pages = {A4},
          doi = {10.1051/0004-6361/201321970},
archivePrefix = {arXiv},
       eprint = {1308.1508},
 primaryClass = {astro-ph.SR},
       adsurl = {https://ui.adsabs.harvard.edu/abs/2013A&A...560A...4R}
}

@ARTICLE{Cegla2016,
       author = {{Cegla}, H.~M. and {Lovis}, C. and {Bourrier}, V. and {Beeck}, B. and {Watson}, C.~A. and {Pepe}, F.},
        title = "{The Rossiter-McLaughlin effect reloaded: Probing the 3D spin-orbit geometry, differential stellar rotation, and the spatially-resolved stellar spectrum of star-planet systems}",
      journal = {\aap},
         year = 2016,
        month = apr,
       volume = {588},
          eid = {A127},
        pages = {A127},
          doi = {10.1051/0004-6361/201527794},
archivePrefix = {arXiv},
       eprint = {1602.00322},
 primaryClass = {astro-ph.EP},
       adsurl = {https://ui.adsabs.harvard.edu/abs/2016A&A...588A.127C}
}

@ARTICLE{Benevolenskaya1999,
       author = {{Benevolenskaya}, E.~E. and {Hoeksema}, J.~T. and {Kosovichev}, A.~G. and {Scherrer}, P.~H.},
        title = "{The Interaction of New and Old Magnetic Fluxes at the Beginning of Solar Cycle 23}",
      journal = {\apjl},
         year = 1999,
        month = jun,
       volume = {517},
       number = {2},
        pages = {L163-L166},
          doi = {10.1086/312046},
archivePrefix = {arXiv},
       eprint = {astro-ph/9903404},
 primaryClass = {astro-ph},
       adsurl = {https://ui.adsabs.harvard.edu/abs/1999ApJ...517L.163B}
}

@ARTICLE{Hritam2024,
       author = {{Chakraborty}, H. and {Lendl}, M. and {Akinsanmi}, B. and {Petit dit de la Roche}, D.~J.~M. and {Deline}, A.},
        title = "{SAGE: A tool for constraining the impacts of stellar activity on transmission spectroscopy}",
      journal = {\aap},
         year = 2024,
        month = may,
       volume = {685},
          eid = {A173},
        pages = {A173},
          doi = {10.1051/0004-6361/202347727},
archivePrefix = {arXiv},
       eprint = {2311.16864},
 primaryClass = {astro-ph.EP},
       adsurl = {https://ui.adsabs.harvard.edu/abs/2024A&A...685A.173C}
}

@ARTICLE{Reinhold2015,
       author = {{Reinhold}, Timo and {Gizon}, Laurent},
        title = "{Rotation, differential rotation, and gyrochronology of active Kepler stars}",
      journal = {\aap},
         year = 2015,
        month = nov,
       volume = {583},
          eid = {A65},
        pages = {A65},
          doi = {10.1051/0004-6361/201526216},
archivePrefix = {arXiv},
       eprint = {1507.07757},
 primaryClass = {astro-ph.SR},
       adsurl = {https://ui.adsabs.harvard.edu/abs/2015A&A...583A..65R}
}

@ARTICLE{Balona2016,
       author = {{Balona}, L.~A. and {Abedigamba}, O.~P.},
        title = "{Differential rotation in K, G, F and A stars}",
      journal = {\mnras},
         year = 2016,
        month = sep,
       volume = {461},
       number = {1},
        pages = {497-506},
          doi = {10.1093/mnras/stw1443},
archivePrefix = {arXiv},
       eprint = {1604.07003},
 primaryClass = {astro-ph.SR},
       adsurl = {https://ui.adsabs.harvard.edu/abs/2016MNRAS.461..497B}
}

@ARTICLE{Reiners2007,
       author = {{Reiners}, A.},
        title = "{Differential rotation in F stars}",
      journal = "{Astron. Nachr.}",
         year = 2007,
        month = dec,
       volume = {328},
       number = {10},
        pages = {1034},
          doi = {10.1002/asna.200710853},
archivePrefix = {arXiv},
       eprint = {0710.2398},
 primaryClass = {astro-ph},
       adsurl = {https://ui.adsabs.harvard.edu/abs/2007AN....328.1034R}
}

@ARTICLE{Gilhool2019,
       author = {{Gilhool}, Steven H. and {Blake}, Cullen H.},
        title = "{A Data-driven Technique for Measuring Stellar Rotation}",
      journal = {\apj},
         year = 2019,
        month = apr,
       volume = {875},
       number = {1},
          eid = {8},
        pages = {8},
          doi = {10.3847/1538-4357/ab0a74},
archivePrefix = {arXiv},
       eprint = {1902.11182},
 primaryClass = {astro-ph.SR},
       adsurl = {https://ui.adsabs.harvard.edu/abs/2019ApJ...875....8G}
}

@ARTICLE{Waite2011,
       author = {{Waite}, I.~A. and {Marsden}, S.~C. and {Carter}, B.~D. and {Al{\'e}cian}, E. and {Brown}, C. and {Burton}, D. and {Hart}, R.},
        title = "{High-resolution Spectroscopy and Spectropolarimetry of Some Late F- / Early G-type Sun-like Stars as Targets for Zeeman Doppler Imaging}",
      journal = {\pasa},
         year = 2011,
        month = nov,
       volume = {28},
       number = {4},
        pages = {323-337},
          doi = {10.1071/AS11025},
archivePrefix = {arXiv},
       eprint = {1109.3278},
 primaryClass = {astro-ph.SR},
       adsurl = {https://ui.adsabs.harvard.edu/abs/2011PASA...28..323W}
}

@ARTICLE{Petit2002,
       author = {{Petit}, P. and {Donati}, J. -F. and {Collier Cameron}, A.},
        title = "{Differential rotation of cool active stars: the case of intermediate rotators}",
      journal = {\mnras},
         year = 2002,
        month = aug,
       volume = {334},
       number = {2},
        pages = {374-382},
          doi = {10.1046/j.1365-8711.2002.05529.x},
       adsurl = {https://ui.adsabs.harvard.edu/abs/2002MNRAS.334..374P}
}

@ARTICLE{Schrijver2001,
       author = {{Schrijver}, Carolus J. and {Title}, Alan M.},
        title = "{On the Formation of Polar Spots in Sun-like Stars}",
      journal = {\apj},
         year = 2001,
        month = apr,
       volume = {551},
       number = {2},
        pages = {1099-1106},
          doi = {10.1086/320237},
       adsurl = {https://ui.adsabs.harvard.edu/abs/2001ApJ...551.1099S}
}

@ARTICLE{Kitchatinov1995,
       author = {{Kitchatinov}, L.~L. and {Rüdiger}, G.},
        title = "{Differential rotation in solar-type stars: revisiting the Taylor-number puzzle.}",
      journal = {\aap},
         year = 1995,
        month = jul,
       volume = {299},
        pages = {446},
       adsurl = {https://ui.adsabs.harvard.edu/abs/1995A&A...299..446K}
}

@ARTICLE{Gallagher2002,
       author = {{Gallagher}, Peter T. and {Moon}, Y. -J. and {Wang}, Haimin},
        title = "{Active-Region Monitoring and Flare Forecasting   I. Data Processing and First Results}",
      journal = {\solphys},
         year = 2002,
        month = sep,
       volume = {209},
       number = {1},
        pages = {171-183},
          doi = {10.1023/A:1020950221179},
       adsurl = {https://ui.adsabs.harvard.edu/abs/2002SoPh..209..171G}
}

@ARTICLE{Solanki2003,
author = {{Solanki}, Sami K.},
title = "{Sunspots: An Overview}",
journal = {\aapr},
year = 2003,
month = jan,
volume = {11},
number = {2-3},
pages = {153-286},
doi = {10.1007/s00159-003-0018-4},
adsurl = {https://ui.adsabs.harvard.edu/abs/2003A&ARv..11..153S}
}

@ARTICLE{Osipova2022,
       author = {{Osipova}, A.~A. and {Nagovitsyn}, Yu. A.},
        title = "{Differential Rotation of Large Long-Lived Sunspot Groups and Their Morphological Structure}",
      journal = {Astron. Lett.},
         year = 2022,
        month = nov,
       volume = {48},
       number = {11},
        pages = {682-687},
          doi = {10.1134/S106377372211010X},
       adsurl = {https://ui.adsabs.harvard.edu/abs/2022AstL...48..682O}
}

@BOOK{Carrington1863,
       author = {{Carrington}, Richard Christopher},
        title = "{Observations of the spots on the Sun: from November 9, 1853, to March 24, 1861, made at Redhill}",
         year = 1863,
       adsurl = {https://ui.adsabs.harvard.edu/abs/1863oss..book.....C}
}

@ARTICLE{Foukal1979,
       author = {{Foukal}, P.},
        title = "{A Doppler Measurement with low Scattered Light of the Higher Rotation Rate of Sunspot Magnetic Fields at the Photosphere.}",
      journal = {\apj},
         year = 1979,
        month = dec,
       volume = {234},
        pages = {716-722},
          doi = {10.1086/157548 },
       adsurl = {https://ui.adsabs.harvard.edu/abs/1979ApJ...234..716F}
}

@ARTICLE{BalthasarWoehl1980,
       author = {{Balthasar}, H. and {Wöhl}, H.},
        title = "{Differential Rotation and Meridional Motions of Sunspots in the years 1940-1968}",
      journal = {\aap},
         year = 1980,
        month = dec,
       volume = {92},
       number = {1-2},
        pages = {111-116},
       adsurl = {https://ui.adsabs.harvard.edu/abs/1980A&A....92..111B}
}

@ARTICLE{Snodgrass1983,
author = {{Snodgrass}, H.~B.},
title = "{Magnetic Rotation of the Solar Photosphere}",
journal = {\apj},
year = 1983,
month = jul,
volume = {270},
pages = {288-299},
doi = {10.1086/161121},
adsurl = {https://ui.adsabs.harvard.edu/abs/1983ApJ...270..288S}
}

@ARTICLE{Snodgrass1984,
       author = {{Snodgrass}, H.~B.},
        title = "{Separation of Large-Scale Photospheric Doppler Patterns}",
      journal = {\solphys},
         year = 1984,
        month = aug,
       volume = {94},
       number = {1},
        pages = {13-31},
          doi = {10.1007/BF00154804},
       adsurl = {https://ui.adsabs.harvard.edu/abs/1984SoPh...94...13S}
}

@ARTICLE{NewtonNunn1951,
       author = {{Newton}, H.~W. and {Nunn}, M.~L.},
        title = "{The Sun's Rotation Derived from Sunspots 1934-1944 and Additional Results}",
      journal = {\mnras},
         year = 1951,
        month = jan,
       volume = {111},
        pages = {413},
          doi = {10.1093/mnras/111.4.413},
       adsurl = {https://ui.adsabs.harvard.edu/abs/1951MNRAS.111..413N}
}

@ARTICLE{HowardGilman1984,
       author = {{Howard}, R. and {Gilman}, P.~I. and {Gilman}, P.~A.},
        title = "{Rotation of the Sun Measured from Mount Wilson White-Light Images}",
      journal = {\apj},
         year = 1984,
        month = aug,
       volume = {283},
        pages = {373-384},
          doi = {10.1086/162315},
       adsurl = {https://ui.adsabs.harvard.edu/abs/1984ApJ...283..373H}
}

@ARTICLE{HowardAdkins1983,
       author = {{Howard}, R. and {Adkins}, J.~M. and {Boyden}, J.~E. and {Cragg}, T.~A. and {Gregory}, T.~S. and {Labonte}, B.~J. and {Padilla}, S.~P. and {Webster}, L.},
        title = "{Solar Rotation Results at Mount-Wilson - Part Four - Results}",
      journal = {\solphys},
         year = 1983,
        month = mar,
       volume = {83},
       number = {2},
        pages = {321-338},
          doi = {10.1007/BF00148283},
       adsurl = {https://ui.adsabs.harvard.edu/abs/1983SoPh...83..321H}
}

@ARTICLE{Howard1984,
       author = {{Howard}, Robert},
        title = "{Solar Rotation}",
      journal = {\araa},
         year = 1984,
        month = jan,
       volume = {22},
        pages = {131-155},
          doi = {10.1146/annurev.aa.22.090184.001023},
       adsurl = {https://ui.adsabs.harvard.edu/abs/1984ARA&A..22..131H}
}

@Article{Schroeter1985,
    author={Schr{\"o}ter, E. H.},
    title={The Solar Differential Rotation: Present Status of Oservations},
    journal={\solphys},
    year={1985},
    month={Oct},
    day={01},
    volume={100},
    number={1},
    pages={141-169},
    issn={1573-093X},
    doi={10.1007/BF00158426},
    adsurl={https://doi.org/10.1007/BF00158426}    
}

@ARTICLE{Pesnell,
       author = {{Pesnell}, W. Dean and {Thompson}, B.~J. and {Chamberlin}, P.~C.},
        title = "{The Solar Dynamics Observatory (SDO)}",
      journal = {\solphys},
         year = 2012,
        month = jan,
       volume = {275},
       number = {1-2},
        pages = {3-15},
          doi = {10.1007/s11207-011-9841-3},
       adsurl = {https://ui.adsabs.harvard.edu/abs/2012SoPh..275....3P}
}

@article{Kutsenko2023,
       author = {{Kutsenko}, Alexander S. and {Abramenko}, Valentina I. and {Litvishko}, Daria V.},
        title = "{The Rotation Rate of Solar Active and Ephemeral Regions - II. Temporal Variations of the Rotation Rates}",
      journal = {\mnras},
         year = 2023,
        month = mar,
       volume = {519},
       number = {4},
        pages = {5315-5323},
          doi = {10.1093/mnras/stac3826},
archivePrefix = {arXiv},
       eprint = {2212.14740},
 primaryClass = {astro-ph.SR},
       adsurl = {https://ui.adsabs.harvard.edu/abs/2023MNRAS.519.5315K}
}

@mastersthesis{Balthasar1979,
  author       = {H. Balthasar},
  title        = {Differentielle Rotation und Meridionale Bewegungen der Sonnenflecken in den Jahren 1940 --1968},
  school       = {Georg-August-Universität Göttingen},
  year         = {1979},
  type         = "{Diploma thesis}",
  address      = {Germany}
}

@article{Thompson_2006,
       author = {{Thompson}, W.~T.},
        title = "{Coordinate Systems for Solar Image Data}",
      journal = {\aap},
         year = 2006,
       volume = {449},
       number = {2},
        pages = {791-803},
          doi = {10.1051/0004-6361:20054262},
       adsurl = {https://ui.adsabs.harvard.edu/abs/2006A&A...449..791T}
}

@book{Waldmeier1955,
title = {Ergebnisse und Probleme der Sonnenforschung},
author = {Waldmeier, Max},
year = {1955},
lccn={56027395},
series={Probleme der kosmischen Physik},
publisher={Geest \& Portig},
address = {Leipzig},
adsurl = {https://ui.adsabs.harvard.edu/abs/1955epds.book.....W/abstract},
}

@ARTICLE{Balthasar2007,
       author = {{Balthasar}, H.},
        title = "{Rotational periodicities in sunspot relative numbers}",
      journal = {\aap},
         year = 2007,
        month = aug,
       volume = {471},
       number = {1},
        pages = {281-287},
          doi = {10.1051/0004-6361:20077475},
       adsurl = {https://ui.adsabs.harvard.edu/abs/2007A&A...471..281B}
}

@ARTICLE{Hale1919,
author = {{Hale}, George E. and {Ellerman}, Ferdinand and {Nicholson}, S.~B. and {Joy}, A.~H.},
title = "{The Magnetic Polarity of Sun-Spots}",
journal = {\apj},
year = 1919,
month = apr,
volume = {49},
pages = {153},
doi = {10.1086/142452},
adsurl = {https://ui.adsabs.harvard.edu/abs/1919ApJ....49..153H}
}

@ARTICLE{Scherrer2012,
       author = {{Scherrer}, P.~H. and {Schou}, J. and {Bush}, R.~I. and {Kosovichev}, A.~G. and {Bogart}, R.~S. and {Hoeksema}, J.~T. and {Liu}, Y. and {Duvall}, T.~L. and {Zhao}, J. and {Title}, A.~M. and {Schrijver}, C.~J. and {Tarbell}, T.~D. and {Tomczyk}, S.},
        title = "{The Helioseismic and Magnetic Imager (HMI) Investigation for the Solar Dynamics Observatory (SDO)}",
      journal = {\solphys},
         year = 2012,
        month = jan,
       volume = {275},
       number = {1-2},
        pages = {207-227},
          doi = {10.1007/s11207-011-9834-2},
       adsurl = {https://ui.adsabs.harvard.edu/abs/2012SoPh..275..207S}
}

@INPROCEEDINGS{2003DS9,
   author = {{Joye}, W.~A. and {Mandel}, E.},
    title = "{New Features of SAOImage DS9}",
booktitle = {Astronomical data analysis software and systems XII},
     year = 2003,
   series = {ASP Conf. Ser.},
   volume = 295,
   editor = {{Payne}, H.~E. and {Jedrzejewski}, R.~I. and {Hook}, R.~N.},
    pages = {489},
   adsurl = {http://adsabs.harvard.edu/abs/2003ASPC..295..489J}
}

@INPROCEEDINGS{ISPy2021,
       author = {{D\'iaz Baso}, C. and {Vissers}, G. and {Calvo}, F. and {Pietrow}, A.~G.~M. and {Yadav}, R. and {de la Cruz Rodr{\'\i}guez}, J. and {Zivadinovic}, L.},
        title = "{ISPy}",
    booktitle = {Zenodo software package},
         year = 2021,
       volume = {56},
        month = oct,
          eid = {5608441},
        pages = {5608441},
          doi = {10.5281/zenodo.5608441},
       adsurl = {https://ui.adsabs.harvard.edu/abs/2021zndo...5608441D}
}

\appendix
\onecolumn

\section{Rotation law coefficients of various studies}
\setlength{\LTcapwidth}{0.93\textwidth} 
\begin{longtable}{|l|l|l|l|c|}
\caption{Coefficients for sidereal differential rotation rates in units of [deg day$^{-1}$] obtained by different methods, including the calculated coefficients for L\"oßnitz~1 and L\"oßnitz~2 with uncertainties, in comparison to other works.}
\label{tab:SolarRoationCoefficients_COMPARISON} \\
\hline
 \rule[-4pt]{0pt}{14pt}& \phantom{00000.}Method & \phantom{000000.}$A$ & \phantom{000000.}$B$ & \phantom{00}$C$ \\ \hline
 \rule[-4pt]{0pt}{14pt}\makecell[l]{Lößnitz three-parameter} & $\alpha$ sunspots & $14.336 \pm 0.003$ & $-2.561 \pm 0.060$ & $-2.04\phantom{00} \pm 0.22$ \\ \hline
 \rule[-4pt]{0pt}{14pt}\makecell[l]{Lößnitz two-parameter} & $\alpha$ sunspots & $14.355 \pm 0.002$ & $-3.078 \pm 0.022$ & -- \\ \hline
 \rule[-4pt]{0pt}{14pt}\makecell[l]{\cite{HowardGilman1984}} & all spots & $14.522 \pm 0.004$ & $-2.840 \pm 0.043$ & --  \\ \hline
 \rule[-4pt]{0pt}{14pt}\makecell[l]{\cite{{Balthasar1986}}} & all spots  & $14.551 \pm 0.006$ & $-2.87\phantom{0} \pm 0.06$  & --  \\ \hline
 \rule[-4pt]{0pt}{14pt}\makecell[l]{\cite{{Balthasar1986}}} & Zürich H- \& J-spots  & $14.33 \pm 0.02$ & $-2.86\phantom{0} \pm 0.08$  & --  \\ \hline
 \rule[-4pt]{0pt}{14pt}\makecell[l]{\cite{Tlatov2024}} & all spots  & $14.56\phantom{0} $ & $-3.094 $  & $0.04\phantom{0}$  \\ \hline   
 \rule[-4pt]{0pt}{14pt}\makecell[l]{\cite{NewtonNunn1951}} & recurrent spots & $14.368 \pm 0.004$ & $-2.69\phantom{0} \pm 0.08$  & --  \\ \hline
 \rule[-4pt]{0pt}{14pt}\makecell[l]{\cite{Snodgrass1984}} & Doppler-shift & $14.050 \pm 0.016$ & $-1.492 \pm 0.005$ & $-2.606 \pm 0.019$ \\ \hline
 \rule[-4pt]{0pt}{14pt}\makecell[l]{\cite{HowardAdkins1983}} & Doppler-shift & $14.193$ & $-1.708$ & $-2.361\phantom{000000.}$ \\ \hline
 \rule[-4pt]{0pt}{14pt}\makecell[l]{\cite{Kutsenko2022}} & ARs \& continuum & $14.69\phantom{0} \pm 0.02$  & $-1.17\phantom{0} \pm 0.51$  & $-4.79\phantom{00} \pm 2.44$  \\ \hline
 \rule[-4pt]{0pt}{14pt}\makecell[l]{\cite{Kutsenko2022}} & ARs \& magnetograms & $14.34\phantom{0} \pm 0.01$  & $-2.87\phantom{0} \pm 0.24 $  & $\phantom{-}0.18\phantom{00} \pm 1.08$ \\ \hline
 \rule[-4pt]{0pt}{14pt}\makecell[l]{\cite{Snodgrass1983}} & magnetograms & $14.307 \pm 0.005$ & $-1.980 \pm 0.064$ & $-2.15\phantom{00} \pm 0.11$ \\ \hline 
 \rule[-4pt]{0pt}{14pt}\makecell[l]{\cite{Hathaway2011}} & magnetograms & $14.437 \pm 0.001$ & $-1.48\phantom{0} \pm 0.01$ & $-2.99\phantom{00} \pm 0.01$ \\ \hline 
 \rule[-4pt]{0pt}{14pt}\makecell[l]{\cite{Howe2000}} & helioseismology (0.98 $R_\odot$) & $14.021 \phantom{0}$  & $-1.735 \phantom{0} $  & $-2.162 \phantom{0000000}$ \\ \hline 
 \rule[-4pt]{0pt}{14pt}\makecell[l]{\cite{SchouHowe2002}} & helioseismology (0.98 $R_\odot$) & $14.296 \pm 0.004$  & $-1.018 \pm 0.021 $  & $-3.58\phantom{00} \pm 0.02$ \\ \hline 
 \rule[-4pt]{0pt}{14pt}\makecell[l]{\cite{Rao2024}} & chromosphere (H$\alpha\pm0.5$ \AA) & $14.690 \pm 0.114$  & $-0.698 \pm 0.391 $  & $-2.88\phantom{00} \pm 0.47$ \\ \hline
\end{longtable}

\section{List of all \boldmath $\alpha$ sunspots between September 2013 and January 2024}
\setlength{\LTcapwidth}{0.68\textwidth} 
\begin{longtable}{|c|c|c|c c|c c|}
\caption{List of all $\alpha$ sunspots used in this study.} \label{tab:all_sunspots} \\
\hline
NOAA & Grade & Latitude & First sighting & Time & Last sighting & Time \rule[-4pt]{0pt}{14pt}\\
\hline
\endfirsthead

\hline
NOAA & Grade & Latitude & First sighting & Time & Last sighting & Time \rule[-4pt]{0pt}{14pt}\\
\hline
\endhead

\hline
\multicolumn{7}{r}{\textit{continued on next page}} \rule{0pt}{10pt}\\
\endfoot

\hline
\endlastfoot

11846 & 3 & $-18^\circ$              & 2013-09-17 & 23:00~UT & 2013-09-28 & 15:00~UT \rule[-4pt]{0pt}{14pt}\\ \hline
11857 & 3 & \phantom{0}$-8^\circ$    & 2013-10-02 & 15:00~UT & 2013-10-13 & 23:00~UT \rule[-4pt]{0pt}{14pt}\\ \hline
11864 & 1 & $-23^\circ$              & 2013-10-07 & 23:00~UT & 2013-10-18 & 22:00~UT \rule[-4pt]{0pt}{14pt}\\ \hline
11896 & 3 & $\phantom{-}11^\circ$    & 2013-11-10 & 19:00~UT & 2013-11-22 & 18:00~UT \rule[-4pt]{0pt}{14pt}\\ \hline
11903 & 2 & $-14^\circ$              & 2013-11-17 & 20:00~UT & 2013-11-29 & 19:00~UT \rule[-4pt]{0pt}{14pt}\\ \hline
11912 & 3 & $-21^\circ$              & 2013-12-02 & 19:00~UT & 2013-12-13 & 20:00~UT \rule[-4pt]{0pt}{14pt}\\ \hline
11931 & 2 & $-14^\circ$              & 2013-12-18 & 20:00~UT & 2013-12-30 & 23:00~UT \rule[-4pt]{0pt}{14pt}\\ \hline
11948 & 1 & $\phantom{-0} 6^\circ$   & 2014-01-05 & 20:00~UT & 2014-01-17 & 19:00~UT \rule[-4pt]{0pt}{14pt}\\ \hline
11952 & 1 & $-31^\circ$              & 2014-01-11 & 19:00~UT & 2014-01-23 & 20:00~UT \rule[-4pt]{0pt}{14pt}\\ \hline
12005 & 2 & $\phantom{-}13^\circ$    & 2014-03-12 & 19:00~UT & 2014-03-24 & 18:00~UT \rule[-4pt]{0pt}{14pt}\\ \hline
12032 & 1 & $\phantom{-}13^\circ$    & 2014-04-07 & 23:00~UT & 2014-04-19 & 08:00~UT \rule[-4pt]{0pt}{14pt}\\ \hline
12033 & 1 & $\phantom{-}12^\circ$    & 2014-04-09 & 18:00~UT & 2014-04-20 & 18:00~UT \rule[-4pt]{0pt}{14pt}\\ \hline
12042 & 0 & $\phantom{-}19^\circ$    & 2014-04-16 & 12:00~UT & 2014-04-27 & 18:00~UT \rule[-4pt]{0pt}{14pt}\\ \hline
12057 & 1 & $\phantom{-}15^\circ$    & 2014-05-07 & 04:00~UT & 2014-05-17 & 23:00~UT \rule[-4pt]{0pt}{14pt}\\ \hline
12061 & 2 & $-24^\circ$              & 2014-05-11 & 09:00~UT & 2014-05-23 & 09:00~UT \rule[-4pt]{0pt}{14pt}\\ \hline
12079 & 1 & $\phantom{-}12^\circ$    & 2014-06-01 & 06:00~UT & 2014-06-11 & 23:00~UT \rule[-4pt]{0pt}{14pt}\\ \hline
12090 & 2 & $\phantom{-}24^\circ$    & 2014-06-10 & 19:00~UT & 2014-06-23 & 10:00~UT \rule[-4pt]{0pt}{14pt}\\ \hline
12150 & 2 & $-14^\circ$              & 2014-08-22 & 23:00~UT & 2014-09-02 & 21:00~UT \rule[-4pt]{0pt}{14pt}\\ \hline
12151 & 2 & \phantom{0}$.7^\circ$    & 2014-08-23 & 22:00~UT & 2014-09-03 & 20:00~UT \rule[-4pt]{0pt}{14pt}\\ \hline
12186 & 2 & $-21^\circ$              & 2014-10-07 & 23:00~UT & 2014-10-19 & 20:00~UT \rule[-4pt]{0pt}{14pt}\\ \hline
12187 & 3 & $-10^\circ$              & 2014-10-11 & 00:00~UT & 2014-10-23 & 23:00~UT \rule[-4pt]{0pt}{14pt}\\ \hline
12195 & 1 & $\phantom{-0} 7^\circ$   & 2014-10-22 & 20:00~UT & 2014-11-02 & 23:00~UT \rule[-4pt]{0pt}{14pt}\\ \hline
12218 & 3 & $\phantom{-}16^\circ$    & 2014-11-23 & 23:00~UT & 2014-12-06 & 16:00~UT \rule[-4pt]{0pt}{14pt}\\ \hline
12246 & 3 & $\phantom{-}18^\circ$    & 2014-12-21 & 21:00~UT & 2015-01-02 & 22:00~UT \rule[-4pt]{0pt}{14pt}\\ \hline
12341 & 2 & $-19^\circ$              & 2015-05-07 & 20:00~UT & 2015-05-19 & 16:00~UT \rule[-4pt]{0pt}{14pt}\\ \hline
12356 & 1 & $-15^\circ$              & 2015-05-27 & 22:00~UT & 2015-01-08 & 04:00~UT \rule[-4pt]{0pt}{14pt}\\ \hline
12375 & 2 & $-11^\circ$              & 2015-06-29 & 18:00~UT & 2015-07-10 & 22:00~UT \rule[-4pt]{0pt}{14pt}\\ \hline
12384 & 1 & $-18^\circ$              & 2015-07-07 & 23:00~UT & 2015-07-19 & 15:00~UT \rule[-4pt]{0pt}{14pt}\\ \hline
12501 & 2 & $\phantom{-0} 4^\circ$   & 2016-02-14 & 21:00~UT & 2016-02-25 & 23:00~UT \rule[-4pt]{0pt}{14pt}\\ \hline
12513 & 3 & $\phantom{-}10^\circ$    & 2016-03-02 & 22:00~UT & 2016-03-13 & 22:00~UT \rule[-4pt]{0pt}{14pt}\\ \hline
12519 & 2 & $\phantom{-0} 6^\circ$   & 2016-03-09 & 21:00~UT & 2016-03-20 & 23:00~UT \rule[-4pt]{0pt}{14pt}\\ \hline
12526 & 2 & \phantom{0}$-4^\circ$    & 2016-03-24 & 18:00~UT & 2016-04-05 & 04:00~UT \rule[-4pt]{0pt}{14pt}\\ \hline
12533 & 3 & \phantom{0}$-2^\circ$    & 2016-04-20 & 18:00~UT & 2016-05-01 & 23:00~UT \rule[-4pt]{0pt}{14pt}\\ \hline
12553 & 3 & \phantom{0}$-7^\circ$    & 2016-06-10 & 06:00~UT & 2016-06-21 & 23:00~UT \rule[-4pt]{0pt}{14pt}\\ \hline
12600 & 3 & $\phantom{-}11^\circ$    & 2016-10-06 & 18:00~UT & 2016-10-18 & 08:00~UT \rule[-4pt]{0pt}{14pt}\\ \hline
12638 & 2 & $\phantom{-}18^\circ$    & 2017-02-19 & 23:00~UT & 2017-03-03 & 12:00~UT \rule[-4pt]{0pt}{14pt}\\ \hline
12662 & 1 & $\phantom{-}12^\circ$    & 2017-06-13 & 06:00~UT & 2017-06-24 & 10:00~UT \rule[-4pt]{0pt}{14pt}\\ \hline
12670 & 2 & \phantom{0}$-6^\circ$    & 2017-08-01 & 22:00~UT & 2017-08-13 & 05:00~UT \rule[-4pt]{0pt}{14pt}\\ \hline
12680 & 2 & $\phantom{-0} 8^\circ$   & 2017-09-09 & 19:00~UT & 2017-09-21 & 19:00~UT \rule[-4pt]{0pt}{14pt}\\ \hline
12681 & 1 & $-14^\circ$              & 2017-09-20 & 18:00~UT & 2017-10-01 & 23:00~UT \rule[-4pt]{0pt}{14pt}\\ \hline
12682 & 2 & $-11^\circ$              & 2017-09-24 & 22:00~UT & 2017-10-05 & 22:00~UT \rule[-4pt]{0pt}{14pt}\\ \hline
12685 & 1 & \phantom{0}$-9^\circ$    & 2017-10-21 & 12:00~UT & 2017-11-01 & 19:00~UT \rule[-4pt]{0pt}{14pt}\\ \hline
12686 & 1 & $\phantom{-}13^\circ$    & 2017-10-23 & 08:00~UT & 2017-11-03 & 19:00~UT \rule[-4pt]{0pt}{14pt}\\ \hline
12738 & 3 & $\phantom{-0} 6^\circ$   & 2019-04-07 & 18:00~UT & 2019-04-18 & 22:00~UT \rule[-4pt]{0pt}{14pt}\\ \hline
12741 & 2 & $\phantom{-0} 6^\circ$   & 2019-05-06 & 21:00~UT & 2019-05-18 & 20:00~UT \rule[-4pt]{0pt}{14pt}\\ \hline
12767 & 3 & $-20^\circ$              & 2020-07-21 & 18:00~UT & 2020-08-02 & 10:00~UT \rule[-4pt]{0pt}{14pt}\\ \hline
12770 & 0 & $\phantom{-}22^\circ$    & 2020-08-04 & 09:00~UT & 2020-08-14 & 18:00~UT \rule[-4pt]{0pt}{14pt}\\ \hline
12776 & 0 & $-14^\circ$              & 2020-10-14 & 23:00~UT & 2020-10-25 & 10:00~UT \rule[-4pt]{0pt}{14pt}\\ \hline
12783 & 2 & $-23^\circ$              & 2020-11-17 & 10:00~UT & 2020-11-28 & 23:00~UT \rule[-4pt]{0pt}{14pt}\\ \hline
12785 & 2 & $-22^\circ$              & 2020-11-22 & 19:00~UT & 2020-12-03 & 23:00~UT \rule[-4pt]{0pt}{14pt}\\ \hline
12790 & 1 & $-23^\circ$              & 2020-11-30 & 20:00~UT & 2020-12-12 & 06:00~UT \rule[-4pt]{0pt}{14pt}\\ \hline
12794 & 3 & $-16^\circ$              & 2020-12-21 & 11:00~UT & 2021-01-01 & 23:00~UT \rule[-4pt]{0pt}{14pt}\\ \hline
12818 & 1 & $-16^\circ$              & 2021-04-19 & 23:00~UT & 2021-05-01 & 18:00~UT \rule[-4pt]{0pt}{14pt}\\ \hline
12833 & 2 & $\phantom{-}24^\circ$    & 2021-06-12 & 23:00~UT & 2021-06-24 & 23:00~UT \rule[-4pt]{0pt}{14pt}\\ \hline
12882 & 1 & $\phantom{-}20^\circ$    & 2021-10-03 & 23:00~UT & 2021-10-15 & 23:00~UT \rule[-4pt]{0pt}{14pt}\\ \hline
12886 & 3 & $-19^\circ$              & 2021-10-17 & 23:00~UT & 2021-10-29 & 18:00~UT \rule[-4pt]{0pt}{14pt}\\ \hline
12893 & 3 & $\phantom{-}17^\circ$    & 2021-10-31 & 00:00~UT & 2021-11-12 & 10:00~UT \rule[-4pt]{0pt}{14pt}\\ \hline
12894 & 1 & $-26^\circ$              & 2021-11-05 & 12:00~UT & 2021-11-16 & 22:00~UT \rule[-4pt]{0pt}{14pt}\\ \hline
12896 & 1 & $-18^\circ$              & 2021-11-14 & 18:00~UT & 2021-11-25 & 20:00~UT \rule[-4pt]{0pt}{14pt}\\ \hline
12897 & 0 & $\phantom{-}17^\circ$    & 2021-11-15 & 23:00~UT & 2021-11-26 & 17:00~UT \rule[-4pt]{0pt}{14pt}\\ \hline
12925 & 0 & $-33^\circ$              & 2022-01-05 & 19:00~UT & 2022-01-16 & 22:00~UT \rule[-4pt]{0pt}{14pt}\\ \hline
12934 & 3 & $-25^\circ$              & 2022-01-19 & 20:00~UT & 2022-02-01 & 18:00~UT \rule[-4pt]{0pt}{14pt}\\ \hline
12954 & 2 & $\phantom{-}17^\circ$    & 2022-02-20 & 19:00~UT & 2022-03-03 & 23:00~UT \rule[-4pt]{0pt}{14pt}\\ \hline
12955 & 2 & $\phantom{-}15^\circ$    & 2022-02-21 & 19:00~UT & 2022-03-04 & 23:00~UT \rule[-4pt]{0pt}{14pt}\\ \hline
12995 & 2 & $\phantom{-}15^\circ$    & 2022-04-19 & 06:00~UT & 2022-04-30 & 23:00~UT \rule[-4pt]{0pt}{14pt}\\ \hline
12999 & 1 & $-20^\circ$              & 2022-04-23 & 22:00~UT & 2022-05-05 & 23:00~UT \rule[-4pt]{0pt}{14pt}\\ \hline
13001 & 0 & $-32^\circ$              & 2022-04-26 & 22:00~UT & 2022-05-08 & 23:00~UT \rule[-4pt]{0pt}{14pt}\\ \hline
13024 & 2 & $-33^\circ$              & 2022-04-25 & 01:00~UT & 2022-06-06 & 12:00~UT \rule[-4pt]{0pt}{14pt}\\ \hline
13034 & 2 & $\phantom{-0} 1^\circ$   & 2022-06-12 & 23:00~UT & 2022-06-23 & 09:00~UT \rule[-4pt]{0pt}{14pt}\\ \hline
13062 & 3 & $-25^\circ$              & 2022-07-19 & 19:00~UT & 2022-07-31 & 22:00~UT \rule[-4pt]{0pt}{14pt}\\ \hline
13071 & 2 & $-19^\circ$              & 2022-08-02 & 21:00~UT & 2022-08-13 & 23:00~UT \rule[-4pt]{0pt}{14pt}\\ \hline
13074 & 1 & $-17^\circ$              & 2022-08-05 & 08:00~UT & 2022-08-16 & 22:00~UT \rule[-4pt]{0pt}{14pt}\\ \hline
13092 & 1 & $-10^\circ$              & 2022-08-31 & 23:00~UT & 2022-09-12 & 21:00~UT \rule[-4pt]{0pt}{14pt}\\ \hline
13111 & 1 & $\phantom{-}28^\circ$    & 2022-09-26 & 23:00~UT & 2022-10-08 & 23:00~UT \rule[-4pt]{0pt}{14pt}\\ \hline
13156 & 1 & $\phantom{-}25^\circ$    & 2022-12-01 & 20:00~UT & 2022-12-12 & 20:00~UT \rule[-4pt]{0pt}{14pt}\\ \hline
13160 & 3 & $\phantom{-}22^\circ$    & 2022-12-06 & 20:00~UT & 2022-12-18 & 12:00~UT \rule[-4pt]{0pt}{14pt}\\ \hline
13162 & 2 & $-13^\circ$              & 2022-12-08 & 12:00~UT & 2022-12-19 & 20:00~UT \rule[-4pt]{0pt}{14pt}\\ \hline
13168 & 3 & $-16^\circ$              & 2022-12-14 & 10:00~UT & 2022-12-26 & 10:00~UT \rule[-4pt]{0pt}{14pt}\\ \hline
13198 & 0 & $\phantom{-}27^\circ$    & 2023-01-18 & 19:00~UT & 2023-01-29 & 19:00~UT \rule[-4pt]{0pt}{14pt}\\ \hline
13201 & 1 & $\phantom{-}23^\circ$    & 2023-01-23 & 20:00~UT & 2023-02-03 & 12:00~UT \rule[-4pt]{0pt}{14pt}\\ \hline
13220 & 2 & $-14^\circ$              & 2023-02-09 & 20:00~UT & 2023-02-21 & 20:00~UT \rule[-4pt]{0pt}{14pt}\\ \hline
13238 & 2 & $\phantom{-0} 9^\circ$   & 2023-02-26 & 19:00~UT & 2023-03-09 & 16:00~UT \rule[-4pt]{0pt}{14pt}\\ \hline
13239 & 3 & $\phantom{-}31^\circ$    & 2023-02-28 & 20:00~UT & 2023-03-11 & 23:00~UT \rule[-4pt]{0pt}{14pt}\\ \hline
13241 & 1 & $\phantom{-}27^\circ$    & 2023-03-01 & 20:00~UT & 2023-03-12 & 20:00~UT \rule[-4pt]{0pt}{14pt}\\ \hline
13251 & 1 & $-13^\circ$              & 2023-03-08 & 23:00~UT & 2023-03-20 & 23:00~UT \rule[-4pt]{0pt}{14pt}\\ \hline
13256 & 1 & $-22^\circ$              & 2023-03-17 & 19:00~UT & 2023-03-29 & 22:00~UT \rule[-4pt]{0pt}{14pt}\\ \hline
13262 & 1 & $-20^\circ$              & 2023-03-22 & 10:00~UT & 2023-04-02 & 23:00~UT \rule[-4pt]{0pt}{14pt}\\ \hline
13264 & 2 & $\phantom{-}15^\circ$    & 2023-03-24 & 12:00~UT & 2023-04-04 & 22:00~UT \rule[-4pt]{0pt}{14pt}\\ \hline
13275 & 1 & $\phantom{-}19^\circ$    & 2023-04-09 & 22:00~UT & 2023-04-20 & 22:00~UT \rule[-4pt]{0pt}{14pt}\\ \hline
13299 & 2 & \phantom{0}$-8^\circ$    & 2023-05-02 & 08:00~UT & 2023-05-14 & 04:00~UT \rule[-4pt]{0pt}{14pt}\\ \hline
13313 & 2 & $\phantom{-}23^\circ$    & 2023-05-19 & 23:00~UT & 2023-05-31 & 18:00~UT \rule[-4pt]{0pt}{14pt}\\ \hline
13320 & 3 & $\phantom{-}10^\circ$    & 2023-06-03 & 06:00~UT & 2023-06-09 & 23:00~UT \rule[-4pt]{0pt}{14pt}\\ \hline
13391 & 2 & $\phantom{-}23^\circ$    & 2023-07-29 & 20:00~UT & 2023-08-10 & 18:00~UT \rule[-4pt]{0pt}{14pt}\\ \hline
13411 & 2 & $\phantom{-}13^\circ$    & 2023-08-15 & 22:00~UT & 2023-08-27 & 22:00~UT \rule[-4pt]{0pt}{14pt}\\ \hline
13412 & 3 & $\phantom{-}30^\circ$    & 2023-08-18 & 06:00~UT & 2023-08-30 & 22:00~UT \rule[-4pt]{0pt}{14pt}\\ \hline
13433 & 2 & $\phantom{-}28^\circ$    & 2023-09-11 & 23:00~UT & 2023-09-23 & 23:00~UT \rule[-4pt]{0pt}{14pt}\\ \hline
13440 & 1 & $\phantom{-}18^\circ$    & 2023-09-18 & 22:00~UT & 2023-09-29 & 23:00~UT \rule[-4pt]{0pt}{14pt}\\ \hline
13448 & 2 & $\phantom{-}13^\circ$    & 2023-09-25 & 20:00~UT & 2023-10-06 & 22:00~UT \rule[-4pt]{0pt}{14pt}\\ \hline
13468 & 1 & $-10^\circ$              & 2023-10-16 & 03:00~UT & 2023-10-27 & 23:00~UT \rule[-4pt]{0pt}{14pt}\\ \hline
13494 & 2 & $-18^\circ$              & 2023-11-20 & 21:00~UT & 2023-12-01 & 23:00~UT \rule[-4pt]{0pt}{14pt}\\ \hline
13501 & 2 & \phantom{0}$-9^\circ$    & 2023-11-23 & 23:00~UT & 2023-11-03 & 23:00~UT \rule[-4pt]{0pt}{14pt}\\ \hline
13505 & 1 & $-17^\circ$              & 2023-11-28 & 16:00~UT & 2023-12-08 & 19:00~UT \rule[-4pt]{0pt}{14pt}\\ \hline
13507 & 2 & $\phantom{-0}8^\circ$    & 2023-11-29 & 14:00~UT & 2023-12-10 & 19:00~UT \rule[-4pt]{0pt}{14pt}\\ \hline
13531 & 1 & $-20^\circ$              & 2023-12-20 & 19:00~UT & 2023-12-31 & 20:00~UT \rule[-4pt]{0pt}{14pt}\\ \hline
13545 & 1 & \phantom{0}$-6^\circ$    & 2024-01-08 & 23:00~UT & 2023-01-20 & 20:00~UT \rule[-4pt]{0pt}{14pt}\\ \hline
\end{longtable}

\end{document}